\documentclass[twocolumn,11pt,final]{report}

\usepackage{fullpage}

\usepackage{showkeys}
\usepackage{amsmath, amssymb, epsf}
\usepackage{amsthm} 
\usepackage{algorithm}
\usepackage{procedure}
\usepackage[noend]{algorithmic}

\usepackage{graphics,graphicx}
\usepackage{subfigure}
\usepackage[pagewise,switch]{lineno}

\usepackage{color}
\usepackage{eso-pic}
\usepackage[24h]{scrtime}

\def\GRIC{ {\bf \texttt{GRIC}} }
\def\GRICplus{ $\textbf{ \texttt{GRIC}}^+$ }
\def\GRICminus{ $\textbf{ \texttt{GRIC}}^-$ }
\newtheorem*{remark}{Remark}
\begin{document}
\title{
{\Huge
Geographic Routing Around Obstacles
in Wireless Sensor Networks
}\\
Technical Report
}
 \author{
    Olivier Powell \thanks{This work was supported by the \emph{Swiss National Science Foundation}, grant number PBGE2 - 112864} \and
    Sotiris Nikoletseas\thanks{
 This work was partially supported by the IST Programme of the European Union, contact 
 number IST-2005-15964 ({\sf AEOLUS}).}
 }
\date{
\{olivier.powell/sotiris.nikoletseas\}@cti.gr
\\
http://tcs.unige.ch/powell
\\
http://www.cti.gr/RD1/nikole/
\\\phantom{aaa}
\Large
Computer Engineering and Informatics Department of Patras University
\\
and
\\
Research Academic Computer Technology Institute (CTI), Greece.
}
\maketitle
\tableofcontents
\begin{abstract}
Geographic routing is becoming the protocol of choice for many sensor
network applications. The current state of the art is unsatisfactory:
some algorithms are very efficient, however they require a preliminary
planarization of the communication graph. Planarization induces
overhead and is not realistic in many scenarios. On the other hand,
georouting algorithms which do not rely on planarization have fairly
low success rates \emph{and} either fail to route messages around all but the
simplest obstacles or have a high topology control overhead (e.g. contour detection algorithms).
To overcome these limitations, we
propose \GRIC \cite{PN07}, the first lightweight and efficient on demand
(i.e. all-to-all) geographic
routing algorithm which does not require planarization \emph{and}
has almost 100\% delivery rates (when no obstacles are added).
Furthermore, the excellent behavior of our algorithm is maintained
even in the presence of large convex obstacles. 
The case of hard concave obstacles is also studied;
such obstacles are hard instances for which performance diminishes.
\end{abstract}
\chapter{Introduction}
We consider the problem of routing in ad hoc wireless networks
of location aware stations, i.e. the stations know their
location in a coordinate system such as the Euclidean plane. 
This problem is commonly called geographic routing (georouting for
short). We are motivated by the case where the network is actually a
wireless sensor network (sensor net) and we address the specific
requirements imposed by their very stringent constraints.
In particular, routing algorithms should be realistic and usable in real world
scenarios, including urban scenario with large communication blocking
obstacle (such as walls, buildings) and areas of low node density
(sometimes called routing holes).
\newtheorem*{problemdf}{Problem definition}
\begin{problemdf}
The problem we are addressing is therefore to find a simple and efficient georouting
algorithm delivering messages with high success rate even in the
presence of large communication blocking obstacles and regions of low
node density.
The georouting algorithms we allow ourselves to consider should be
lightweight, on demand, efficient and realistic.
\end{problemdf}
\emph{Lightweight} is notably understood in terms of topology maintenance
overhead (our algorithm actually requires \emph{no link-layer topology maintenance}),
\emph{on demand} means it should be all-to-all
(as opposed to all-to-one and one-to-all algorithms) and not rely on
storing any routing table, 
\emph{efficiency} is measured in terms of success rate (the
probability that a message reaches its destination) and hop-stretch
(i.e. the path length should be short) and it should be
\emph{realistic} and of practical use for real world sensor nets, thus
avoid relying on any unrealistic assumptions (such as unit disc graphs).
\section{Background on Sensor Networks and Geographic Routing}
Recent advances in micro-electromechanical systems (MEMS) and wireless
networking technologies have enabled the development of very small
sensing devices called sensor nodes \cite{RAS+00,WLL01,ASS+02}.
Unlike traditional sensors that operate in a passive mode, sensor
nodes are smart devices with sensing, data-processing and
wireless transmission capabilities. Data can thus be collected,
processed and shared with neighbours. 
Sensor nodes are meant to be deployed in large \emph{wireless
sensor networks} (sensor nets) instrumenting the physical world.
Originally developed for tactical surveillance in military
applications, their are expected to find growing importance in civil
applications such as building, industrial and home automation,
infrastructure monitoring, agriculture and security. 
Their range of applications makes sensor nets
heterogeneous in terms of size, density and hardware
capacity. 
Their great attractiveness follows the
availability of low cost, low power and miniaturized sensor nodes
pervading, instrumenting  and reality augmenting the physical world
\emph{without requiring human intervention}: sensor nets are
self-organizing, self repairing and highly autonomous.
Those properties are implemented by designing protocols addressing the
specific requirements of sensor nets. Common to all sensor net
architecture and applications is the need to propagate data inside the
network. Almost every scenario implies multi-hop data transmission
because of low power used for radio transmission mostly to limit
depletion of the scare energy resources of battery powered sensor
nodes. 
As a consequence, routing comes
as a fundamental primitive of every full fledged sensor network
protocol, whatever the application, the topology and the hardware
composing the sensor net. Zhao and Guibas \cite{ZG04c}
claim with purpose that routing protocols for sensor
nets need to respond to specific constraints.
\begin{quote}\emph{
``The most appropriate protocols \emph{[for sensor nets]} are those that discover
routes on demand using local, lightweight, scalable techniques, while
avoiding the overhead of storing routing tables or other information
that is expensive to update such as link costs or topology
changes''}.
\end{quote}
Geographic routing is a very attractive solution for sensor nets.
The most basic georouting algorithm is the greedy georouting
algorithm described by Finn \cite{F87}, where messages are
propagated with minimization of the remaining distance
to the message's destination used as a heuristic for choosing the
next hop relay node. Foundational (and some more advanced) georouting
techniques and limitations are covered in most text books on sensor
nets, e.g. \cite{KW05c,ZG04c}. The greedy
routing algorithm discovers routes on demand and is
all-to-all (as opposed to all-to-one), it uses only location
information of neighbour nodes, it scales perfectly well,
requires no routing tables and adapts dynamically to topology changes.
The only requirement, which is a strong one, is that sensor nodes
have to be
\emph{localized}, i.e. they should have access to coordinates defining
their position. 
The major problem of the greedy approach is that messages are very
likely to get trapped inside of local minimums (nodes who have no
neighbours closer to the destination than themselves). For this reason
early georouting has received much attention and research,
motivated by the gain in momentum of ad hoc wireless networking has
been fruitful in inventing ingenious new georouting techniques which
we review in more details in section \ref{related work}.
\section{Localization}
How reasonable is it to assume that sensor nodes are localized?
It is usually accepted that the use of GPS like technology on every
node is not reasonable for sensor nets. Therefore, one
may ask to what extent the strong hypothesis of localised nodes
restricts the application of georouting techniques to a specialized
niche of sensor nets. First of all, clearly localization of nodes
comes at a cost, but this cost can be kept reasonable through the use
of one of the localization techniques for sensor nets often using GPS
like technology to localize a few beacon
nodes followed by a distributed protocol to localise the bulk of
sensor nodes as described in general books on sensor nets such as
\cite{KW05c,B05,ZG04c}, recent papers such as \cite{LM+06} 
or in the extensive survey in \cite{HB01}.  
Those techniques come at the price of an overhead
in terms of protocol complexity which would diminish the
attractiveness of georouting if it was not for the fact that in many
application localization will be required anyway since, 
as explained by Karl and Willig \cite{KW05c}:
\begin{quote}\emph{
``...in many circumstances, it is useful and even necessary for a node in a
wireless sensor network to be aware of its location in the physical
world. For example, tracking or event-detection functions are not
particularly useful if the \emph{[sensor net]} cannot provide any information
\emph{where} an event has happened''}.
\end{quote}
As a consequence, georouting does not confine to a
specialised niche but on the contrary it is becoming a protocol of
choice for sensor nets as advocated by Seada, Helmy and Govindan \cite{SHG04}:
\begin{quote}
``...\emph{\emph{[georouting]} is becoming the protocol of choice for many
emerging applications in sensor networks, such as data-centric
storage \cite{RKY02} and distributed indexing
\cite{GEG03}''}.
\end{quote}
We are therefore confident that georouting will be a sustainable
approach for many sensor net applications and that the ongoing
research will keep improving the state of the art to a point where
localisation will be made available with high precision at low
protocol overhead for sensor nets.
Interestingly, it may be noticed that georouting can even be used
when nodes are not localised by using \emph{virtual coordinates} as
was proposed in \cite{RRP03}.
\section{Our approach}
To overcome the limitations of previous approaches, we propose a
new algorithm called \GRIC\footnote{\GRIC is pronounced ``Greek'' in
reference to the fact that it has been designed in the University of
Patras in Greece.} (\textbf{G}eo{\bf R}out{\bf I}ng around
obsta{\bf C}les) \cite{PN07}. The main idea of \GRIC is to appropriately combine
movement directly towards the  destination to optimize performance
with an inertia effect that forces messages to keep moving along the
``current'' direction so that it will closely follow obstacle shapes in order to
efficiently bypass them. In order to route messages around more complex
obstacles we add a ``right-hand rule'' inspired component to our
algorithm, which is borrowed from maze solving algorithms (algorithms
one can use to find he's way out of a maze).
The right-hand rule is a well known ``wall
follower'' technique to get out of a maze \cite{H89} which has been previously
used by some of the most successful georouting algorithms known so far (c.f. section
\ref{related work}). However, whereas previous algorithms used a
strict implementation of the right hand rule we only inspire ourselves
from it. The reason is that the right-hand rule, originally developed to
find a path out of a maze, only works for planar graphs but we want to
ensure that \GRIC runs on the complete (non
planarized) communication graph.
\subsection{Methodology}
We implement our algorithm and comparatively evaluate its performance
against those of other representative algorithms (greedy, LTP, and
face routing).
We focus on three performance measures: success rate, distance
travelled and hop count. We study the impact on performance of several
types of obstacles (both convex and concave) and representative
regimes of network density.  
\subsection{Comparison}
When compared to other georouting algorithms,
\GRIC features no topology maintenance overhead, practical usefulness
in the sense that it runs
on  real world complete (directed) communication graphs
and is capable of routing messages to their destination with high
success rate,  even in the presence of large communication blocking
obstacles. The combination of those features makes \GRIC unique. 
Some previous algorithms had high success rates, however none of them works
on directed communication graphs and they all suffer either from a
high topology maintenance overhead or are not practical because
requiring the unrealistic assumption that the communication graph is
(almost) a unit disc graph. Other algorithms are lightweight (i.e. have
low topology maintenance overhead), but none of them has competing
success rates when compared to \GRIC, furthermore none of them is capable of routing
messages around large obstacles.
\subsection{Strengths of our approach:}
Our algorithm has the advantage of
being many-to-many and on demand, i.e. no routing tables have to be
stored, no flooding is required to establish a gradient and no interests
have to be propagated for the routing to be successful. This is possible
because the nodes are localised and arguably is part of the strengths
inherent to all georouting algorithms. It is in large part responsible
for the attractiveness of this kind of protocols.

Our algorithm seems to somehow
combine the best of two worlds. \emph{In terms of algorithm design}, 
it combines the light weight and simplicity of
the greedy type of algorithms while avoiding the (sometimes
unrealistic and high topology maintenance overhead) topology maintenance
phase (usually planarization) required by the GFG \cite{BMS99} and GPSR \cite{KK00} algorithms
and their successors.
c.f. section \ref{issues} for a detailed
explanation of the limitations induced by the planarization phase.
\emph{In terms of performance}, our findings demonstrate that in most
cases our algorithm outperforms the previous ones. Furthermore, its
performance with respect to the path length is very close to the
length of an optimal path in the case of no global knowledge,
e.g. when the presence of obstacles is not known to the algorithm in
advance but only sensed when reaching the obstacles. 
It may also be worth pointing out that our algorithm seems to be
resistant to temporary link failure, as  suggested by the fact that
the randomized version of \texttt{GRIC} actually performs better than
the deterministic one, c.f. section \ref{randomization}
Because our algorithm is the first lightweight algorithm to
efficiently bypass large and complex obstacles without requiring a
preliminary planarization phase, we feel it \emph{considerably advances the
current state of the art} for gerouting algorithms.
\section{Related Work}
\label{related work}
The major problem of the early greedy geometric routing algorithms \cite{F87} is the
so called \emph{routing hole problem}
\cite{KW05c,ZG04c,AKJ05} 
where messages get trapped by ``local
minimum'' nodes which have no neighbours closer to the destination of
the message than themselves. The incidence of routing holes
increases as network density diminishes and the success rate of the
greedy algorithm
drops very quickly with network density.
So far, georouting protocol have mainly focused on
increasing the success rate (the probability that a message is
successfully routed) while keeping the ``lightweight''
advantages of the greedy type of routing algorithms, although recent
research efforts have also been proposed to deliver efficient
georouting algorithms with respect to other metrics \cite{SD04,KNS06}.
Most georouting algorithms can be divided into two groups. On the one hand, those
that \emph{guarantee delivery} (i.e. the success rate is 100\% if the
communication graph is connected) \cite{BMS99,KK00} and \emph{probabilistic
approaches} that achieve certain performance trade-offs, such as \cite{CDN04,CDN06,CNS02,CNS05,KNS06}.
Alternatively, it is also possible to run topology discovery
algorithms to discover routing holes and obstacles and to use this
information to route messages around them. An example is
the BOUNDHOLE algorithm from \cite{FGG06} which uses the TENT rule to
discover local minimum nodes and then to ``bound'' the
contour of routing holes. The disadvantage of this type of approach
is the heavy overhead and the fragility towards unstable topology
(such as mobility of nodes). Although it has a very high overhead when
compared to the algorithm we propose,
the information gained during the contour discovery phase may be used
for other application than routing such as path migration,
information storage mechanisms and identification of regions of
interest \cite{FGG06}.  
\section{Routing with Guaranteed Delivery}
Arguably, most georouting algorithms are inspired by the routing
protocols with guaranteed delivery from \cite{BMS99,KK00}.
Those two protocols are almost similar and were, to our knowledge,
developed independently. They consist of using a greedy
routing algorithm until the message gets trapped in a routing
hole. When the message gets trapped in the routing hole, the algorithm
switches to a \emph{rescue mode} until it is judged that switching
back to greedy mode is appropriate. The rescue mode
routes messages according to the face algorithm described by Prosenjit
et al. \cite{BMS99} on a planar subgraph of the communication
graph. The face algorithm improves on the compass-II routing
algorithm from \cite{KSU99}: the length of the path along which
messages are routed by face is smaller than for
compass-II\footnote{Compass-II should not be mistaken with compass-I,
  which was also described in \cite{KSU99} and guarantees delivery on
  the Delaunay triangulation of a planar graph. It seems
  \cite{KSU99} were mainly interested in finding conditions under
  which greedy geometric routing succeeds while minimizing the angle
  (and not the distance) to the destination. This is what their
  compass-I (simply called compass) algorithm does. Probably they did not realise the potential
  applications of their compass-II algorithm, when used in combination
  with a distributed planarization algorithm such as those used in
  \cite{BMS99,KK00}.}.
In \cite{BMS99} the FACE algorithm is used on the Gabriel
subgraph of the complete communication graph, whereas in
\cite{KK00} the comparable relative neighbourhood graph is used
\footnote{Both can be
  computed by a distributed algorithm, although the Gabriel graph is
  arguably better since the relative neighbourhood graph is sparser than the
  Gabriel graph (for a same given initial graph).}. Further work proposes to optimize the switching
mechanism between greedy and rescue mode  \cite{KWZ03,KWZ03b} in order
to diminish the length of the routing paths. Face routing, is very
attractive because it guarantees delivery, 
however it involves a number of issues.
The most important issue is the fact that it requires a preliminary
\emph{planarization} phase, a procedure which may not be achievable in
many realistic scenarios. Other issues include
\emph{topology maintenance overhead}, \emph{strong dependency on the unit disc
  graph model}, \emph{sensitivity to localisation errors},
\emph{increased path length}, \emph{requirement of symmetric
  links} and \emph{limitation to 2-dimensional networks}.
Because they motivate our present work, 
we include in section \ref{issues} a detailed comment
on those issues, as well as some of the ways they have been
(or not) addressed in previous works.
\subsection{Issues with face routing}
\label{issues}
First of all, FACE has a clear topology maintenance overhead since
the Gabriel graph has to be maintained at all times. This overhead
is computationally reasonable since the Gabriel graph can be computed by a fully
distributed algorithm \cite{BMS99}. 
A critical issue
is the fact that, as acknowledged in \cite{BFN03}, 
\emph{...the Gabriel graph that is used for routing is not a robust
structure}\footnote{The same holds for the relative neighbourhood
  graph used in \cite{KK00}.}. The problem is that the FACE family of protocols
assume that the communication graph is a random geometric graph built using the unit disc graph
model. Whereas the unit disc graph model is commonly used to model wireless
sensor networks for simulation purposes, 
the fact that it only \emph{approximates} real world communication
graphs for sensor networks
is usually not a critical issue. On the contrary, the correctness of
the FACE algorithm requires the communication graph of a real sensor network to be
\emph{very close} to a unit disc graph. This hypothesis may be very
hard to match in practise and as shown in the micro-analysis of
\cite{SHG04} it results in non-recoverable errors for FACE like
algorithms. 
One of the reason is that the hypothesis
that sensor nodes can communicate to neighbour nodes if and only if
their distance is beyond a given threshold is not true in
real networks. See, for example \cite{ZHK04} for a study of radio
irregularity in sensor networks. In \cite{BFN03}, it is proposed to
introduce virtual edges to overcome this problem. At the cost of
introducing a little more topology maintenance overhead, they show
that FACE is still usable in the case where the communication graph is
a disc graph defined using two radius $r$ and $R$ (a ``two radius''
unit disc graph) with $r\leq R$
such that the ratio $\frac{R}{r}$ is smaller than $\sqrt{2}$. In the
communication graph considered by \cite{BFN03} sensor nodes can always
communicate to neighbours when their distance is smaller than $r$, can
never communicate when their distance is greater than $R$, and
communication is uncertain when the distance between them is between
$r$ and $R$. Although this model is a generalisation of the unit disc
graph model and the findings of \cite{BFN03} shows limited robustness of the
Gabriel graph extraction algorithm permitting successful use of FACE, it is
unclear how the two radius unit disc graph
hypothesis will be matched in real sensor networks, although it will
probably not be the  case in many scenarios, as shown by recent work,
c.f. section \ref{new stuff}.

Yet another
problem follows from localization imprecision of the nodes.
\cite{SHG04} show that even small localisation errors of $10\%$ of the
transmission range have a great impact on the success rate of GPSR,
which is a combination of greedy and FACE routing (on the contrary, it
is shown in \cite{HHB03} that greedy is less affected by this
problem). 
After acknowledging the problem, \cite{SHG04} proposes a fix, called
the \emph{mutual witness} extension to GPSR, which seems to restore almost perfectly the excellent behavior of GPSR that is known to hold under the idealistic
assumption that nodes have access to exact location information.\footnote{The fix is using
the observation that errors mostly follow from the fact that
localisation errors disconnect the Gabriel graph, which is therefore a
problem only when running in rescue mode (since the greedy mode may
run on the full communication graph). Unfortunately, recent findings
from  Kim et al. \cite{pitfalls} show that this fix does not work
well for real world sensor nets (because radio communication graphs
are too different from the theoretical unit disc graph models and the pathologies they
imply on the planarization phases brake face routing ). Also, it may
be noticed that in the case of dense networks,
GPSR mostly runs in greedy mode. It could be that the fix proposed by
the authors does not work well in the case of dense networks if using
the FACE algorithm instead of GPSR, even when the unit disc graph
assumption is not (too strongly) violated. This could be an issue in the case
of running GPSR in a dense network with obstacles such as those we
consider in section \ref{our algo}.} 

Furthermore, as acknowledged in \cite{SD04}, the FACE algorithm runs
on a planar subgraph who's construction favors short edges over long
ones. The path length, which is a common measure of quality for
routing algorithms\footnote{It captures measures such as collision probability,
  since more transmissions imply more risks of loss, delivery delay
  and energy consumption, at least if the transmission power is
  fixed.} 
is thus fairly long for the FACE algorithm. Early fall-back to 
greedy mode \cite{BMS99,KK00,KWZ03,KWZ03b} is one solution to
this problem\footnote{However, in the case where large obstacles are present
  in the network or when the network is dense, a significant part of
  the routing could have to use FACE mode.}. Another solution is to use
``shortcuts'', introduced in \cite{DSW04} to reduce the path length of FACE.
Obviously all those techniques are employed at the cost of an extra
overhead in terms of topology maintenance of the communication
graph.

Another limitation of FACE 
is that it will totally fail in the case where communication links are
asymmetric. This could be a severe limitation for some otherwise
simple and efficient energy aware optimisation schemes for routing algorithms
(e.g. temporarily removing links to nodes with low energy and
favoring links leading to motes with high remaining energy would create
an asymmetric communication graph even if the original communication
graph was symmetric). Even some otherwise efficient energy aware optimisation schemes may not be applicable to FACE, more subtle (and more complicated) approaches are still possible, like the proposition in \cite{SD04} to favor next hop transmission to motes with high available energy in
combination with the ``shortcut'' idea from \cite{DSW04}. Roughly
speaking, instead of choosing the shortcut that minimises the path
length, a shortcut leading to a node with high remaining energy is
chosen in combination with the use of a dominating subset of nodes to
act as a backbone network for routing. 

Finally, it may be noticed that
FACE georouting is not applicable in the case of 3-dimensional networks,
which may be a limitation for some applications of sensor networks. 
\subsection{Recent Progress Towards Realistic Guaranteed Delivery}
\label{new stuff}
As explained in the previous section, until recently the state of the
art for georouting had a major flaw: it strongly depended on the
assumption of a unit disc communication graph, thus making in
unrealistic for many real world scenario. Very recently, Kim et
Al. \cite{pitfalls} acknowledged this fact by presenting a series of pitfalls associated
with geographic routing which extends and deepens the micro-analysis
of \cite{SHG04}. In a breakthrough paper \cite{practical}, the same
authors propose the cross-link detection protocol (CLDP), the first face routing algorithm that works for
arbitrary communication graphs, thus making geographic face routing
\emph{practical} in the sense that their algorithm works for real world
wireless networks whose communication graphs do not need to resemble
idealistic unit disc communication graphs. In fact, the algorithm
works for any graph as long as the links are bi-directional, i.e. the
communication graph is symmetric. The major drawback is that their
solution, while becoming practical for real-world nets, induces a very
high topology maintenance overhead in terms of message exchange in
order to construct a subgraph of the complete communication graph. 
In order to reduce the communication
overhead, Leong et Al. propose \emph{Greedy Distributed Spanning Tree Routing} (GDTSR), a georouting algorithm \cite{no_planar} reducing the 
communication overhead by orders of magnitudes. Their contribution has
the originality of considering a new way of routing messages while in
rescue mode, i.e. as a replacement to face routing. They propose to
use convex-hull spanning trees. Although elegant, their solution
requires building a few complex tree structures. To be made completely
practical their solution would requires handling many tedious problems
(such as selecting the roots of the trees or making the algorithm robust
in case some nodes fails during the set-up phase). Therefore, their
algorithm is far from being simple (it requires the study of
subproblems and implementing solutions to those problems). Those
difficulties affect the robustness of a complete protocol stack (MAC/data-link)
implementation of their algorithm and even if those problems were being
addressed the message overhead, although being reduced when compared
to the solution in \cite{practical}, is still far from being
insignificant. As a sign that those contributions are probably going
to open a new area of creative improvements to georouting, Kim et
Al. recently deepened \cite{lazy} even further the understanding of the pitfalls
of face georouting and, using a lazy cross-link removal (LCR)
algorithm 
propose a practical face routing algorithm with
significantly lower topology maintenance overhead than their previous
proposition from \cite{practical}. Nevertheless, the overhead is still
significant, bi-directional links are still required and the knowledge
of two hops away neighbours is required, which may be a limitation in
case of highly dynamic networks.
\section{Probabilistic approaches:}
Guaranteed delivery algorithms have received plenty of attention in the
literature because, when applicable, they are very powerful
algorithms. 
A taxonomy of georouting algorithms is provided in \cite{S02}, focusing on
guaranteed delivery but with a review of many other solutions.
Some other solutions include the PFR \cite{CDN04,CDN06}, VTRP
\cite{ACM04}, the LTP protocols \cite{CNS02,CNS05} or the CKN protocol from
\cite{CKN06}. 
Those protocols employ randomization and do not guarantee delivery, however they are more
``lightweight'' than the FACE type of algorithms and may also apply in
a more general context than localised network. What is more, they do not
require a planarization phase: they run on the full communication
graph. 
PFR is a probabilistically
limited flooding algorithm (it is thus  multi-path), VTRP proposes to
use variable transmission range capable hardware to bypass routing
holes in low density areas and LTP is a
combination of a modified greedy algorithm with limited backtracking as a
rescue mode. The effect of obstacles on those three protocols was studied in
\cite{CMN06} and therefore provides a comparison to the \GRIC algorithm
proposed in this report. One of the results in \cite{CMN06} is that all
three protocols fail in the case of the stripe obstacle, which is the
easiest obstacle considered in this report. In terms of obstacle
avoidance, \GRIC thus prevails over PFR, VTRP and LTP (as well as over
all other ``lightweight'' georouting algorithms we know of). In the case of no obstacle,
we shall show that \GRIC is more efficient than LTP (and therefore than PFR, as
follows from results in \cite{CMN06}). VTRP is more efficient than \GRIC
in the case of networks of very low densities with no obstacles, but
this is because VTRP uses additional hardware properties, namely, it
``jumps over'' routing holes by augmenting the transmission range
when faced with a blockage situation. Although very nice when
applicable, it may be noticed that this solution requires
additional hardware capabilities, makes the protocol more complicated
(since the MAC and link layer protocol have to adapt, for example
neighbourhood discovery may have to be restarted for the new
transmission range) and does not work if the obstacle is ``physical'',
in the sense that transmission range is blocked (i.e. it only deals
with areas of low node density: routing holes).
More recently, \cite{CKN06}
proposes a routing protocol that implicitly copes with  simple obstacles and
node failures by ``learning'', i.e. gaining limited local knowledge of
the actual network conditions, and by using this information to optimize data
propagation. By planing routes a few hops
ahead and by varying the transmission range it can therefore bypass
routing holes (i.e. areas of low density but through which radio
transmission is possible if the transmission range is increased) 
and some very simple obstacles. However, increasing the transmission
range does not bring any improvement in the case of physical
obstacles (through which radio propagation is impossible) or large
obstacles which can not be ``jumped over'' by increasing the
transmission range.
\chapter{The \GRIC Algorithm}
\section{Sensor Network Model}
\label{sec: SN Model}
This section describes the sensor net model we assume in this work.
The assumptions we make are fairly weak and general. 
We consider sensor networks  deployed on a two
dimensional surface and we allow messages to piggy-back $\mathcal{O}(1)$ bits of
information which is being used for the decision making of our routing
algorithm (in fact, the only required information is the position of
the last node visited by the message, the position of the final
destination of the message and the value of a mark-up flag,
c.f. section \ref{our algo}).
The most important part of the model is the communication model.
We assume that each node is aware of its own position,
has access to the list of its \emph{outbound} neighbours and to their positions,
and that it can reliably send messages to its neighbours, i.e. \GRIC runs
on a directed communication graph where nodes are localised and aware
of their neighbours as well as their positions.
This combinatorial model is reasonable and realistic for sensor nets.
Indeed, our \GRIC algorithm is a \emph{network layer} algorithm.
The network layer relies on an underlying \emph{data link} layer which in
turn relies on a MAC and physical layer.
Real world sensor networks implementation of our algorithm would therefore
have to implement all levels of the protocol stack, and it is
reasonable to assume that the link layer provides the level of
abstraction required by our algorithm. We discuss in more details a
realistic and simple full layer implementation of our algorithm (including
physical, MAC and data link layers) in the next section.
\section{Lower Layers of the Protocol Stack}
For completeness we discuss the assumptions we
make for every layer upon which the network layer relies: the physical, the MAC and the
link layers.

Because our algorithm situates itself at the network
layer level, it is not strongly dependant on the physical layer,
i.e. the type of nodes. 
Highly limited piconodes forming very large smart dust nets
\cite{ASS+02,WLL01,EGH+99} composed of
nodes smaller than a cubic centimeter, weighting less than 100 grammes and with costs well
under a dollar such as envisioned in \cite{RAS+00,WLL01} but with
probably strong limitations in terms of resource and features are perfectly
acceptable. On the other hand more reasonable sensor nodes proposed by
OEM nowadays are fine too.
Those nodes are typically the size of a matchbox (including battery)
and with a cost of more than a hundred dollars.
We assume that each sensor node is a fully-autonomous computing and
communicating device, equipped with a set of sensors (e.g for
temperature, motion, or metal detector). It has a small CPU, some
limited memory and is strongly dependant on its limited available
energy (battery). Each sensor can communicate wireless with
other sensors of the network which are within communication range. 
We do not assume that sensor nodes can vary their transmission range.

We assume that a MAC layer protocol such as the S-MAC
protocol from \cite{YHE02} or an IEEE 802.11-like MAC protocol is
running on each sensor node. The MAC layer protocol is in charge of
neighbourhood discovery and it
provides the link layer protocol with a list of neighbours 
with which the sensor node is capable of symmetric communication.

At the link layer, we assume a protocol that provides each node with a list of reliable
\emph{outbound} wireless communication links (i.e. at the link layer
we do not need to assume the communication link to be symmetric, which
is a feature of our algorithm). This means that the link layer
protocol can remove some of the links provided at the MAC layer
(e.g. those links which cannot be made reliable at reasonable costs,
for example by including some quality metric on the links).

Georouting algorithms all make a further assumption: sensor
nodes are localised for example using one of the currently available localisation
techniques 
\cite{B05,ZG04c,KW05c,HB01}. 
We
assume that the link layer protocol adds this information to the list of
links: the position of the sensor to which messages are sent if a
given wireless link is chosen.

Our \GRIC routing protocol situates itself at the network layer. It
runs on each individual node and the input to the algorithm is the
position of the node on which it runs, as well as the list of links
(including the position of the node to which those links lead to)
provided by the link layer protocol. 

\GRIC is  assumed to be capable
of reliably sending messages to any of its outbound neighbour. The
reliability issue (collision avoidance, acknowledgments, etc) is
not taken into account by \GRIC but by the lower layer protocols (link layer and MAC layer).
Reliable transmission can be achieved through many possible
techniques, like time division multiple access schemes (TDMA) at the
MAC layer,
acknowledgement with retransmission or error correcting codes at the
data-link layer or, alternatively multipath redundancy could be added
on top of \texttt{GRIC}. Because many different options exist, our
assumption of reliable communication links is realistic but out of the
scope of the current report.

\section{Algorithms}
\label{our algo}
Like the FACE family of algorithms and their successors
\cite{BMS99,KK00,KWZ03,KWZ03b}, 
\GRIC ({\bf G}eo{\bf R}out{\bf I}ng around obsta{\bf C}les) \cite{PN07}
uses two different routing modes: a normal mode called \emph{inertia mode}
and a rescue mode called \emph{contour mode}. 
The inertia mode is used when the
message makes progress towards the destination, and the
contour mode when it is going away from the destination. 
The two major ingredients which make our approach successful is the use
of the \emph{inertia principle} and the use of the so called \emph{right-hand rule}.
The inertia principle is borrowed to physics and we use it to control the
trajectory of messages inside the network\footnote{Our combinatorial
   model is not physical, in the sense that our inertia component does
 not follow the rules of physics. We only use physical inertia as an
 inspiration to derive a simple combinatorial rule which mimics
 physical inertia.}. 
Informally, messages are
``attracted'' to their destination (like a celestial body is attracted in a
planet system) but have some inertia (i.e. they also have an incentive
to follow the ``straight line''). The right-hand rule is a ``wall
follower'' technique to get out of a maze \cite{H89} and we use it to
route messages around complex obstacles. 
We use this informal idea to ensure
that messages follow the contour of obstacles. The successful implementation of
both of these abstract principles uses a virtual
\emph{compass device} (described in section \ref{route around obstacles}) which
treats a message's destination as the
north pole. The \emph{compass device}, the \emph{inertia
principle} and the \emph{right-hand rule} enables the computation of an
idealistic direction in which the message should be sent (c.f. figure
\ref{fig compass}). 
Since there may not be a node in this exact direction, the message is actually
sent to the mote maximizing progress towards the ideal direction.

\subsection{Routing with inertia:}
\label{sec inertia}
In this section, we give a high level description of the \emph{inertia routing mode}. 
Intuitively, the inertia mode performs \emph{greedy routing}
with an adjunct \emph{inertia conservation} parameter
$\beta$ ranging in $\left[ 0, 1 \right]$. 
When a node routes a message in \emph{inertia routing mode}, 
it starts by
computing an \emph{ideal direction vector} $v_{\textrm{ideal}}$.
Once this ideal direction vector is computed,
the message is sent to the node maximizing progression in the ideal
direction. We next describe how to compute $v_{\textrm{ideal}}$ and
how to choose the node with best progression towards the ideal direction.
To compute the ideal direction, the routing node needs to know its own position $p$,
the position $p'$ of the node from which the message was received
and the position $p''$ of the message's destination, c.f. figure
\ref{fig compass1} below.
\begin{figure*}[htb]
\centering
\subfigure[Ideal direction.]{
\label{fig compass1}
\includegraphics[angle=0,width=.55\textwidth]{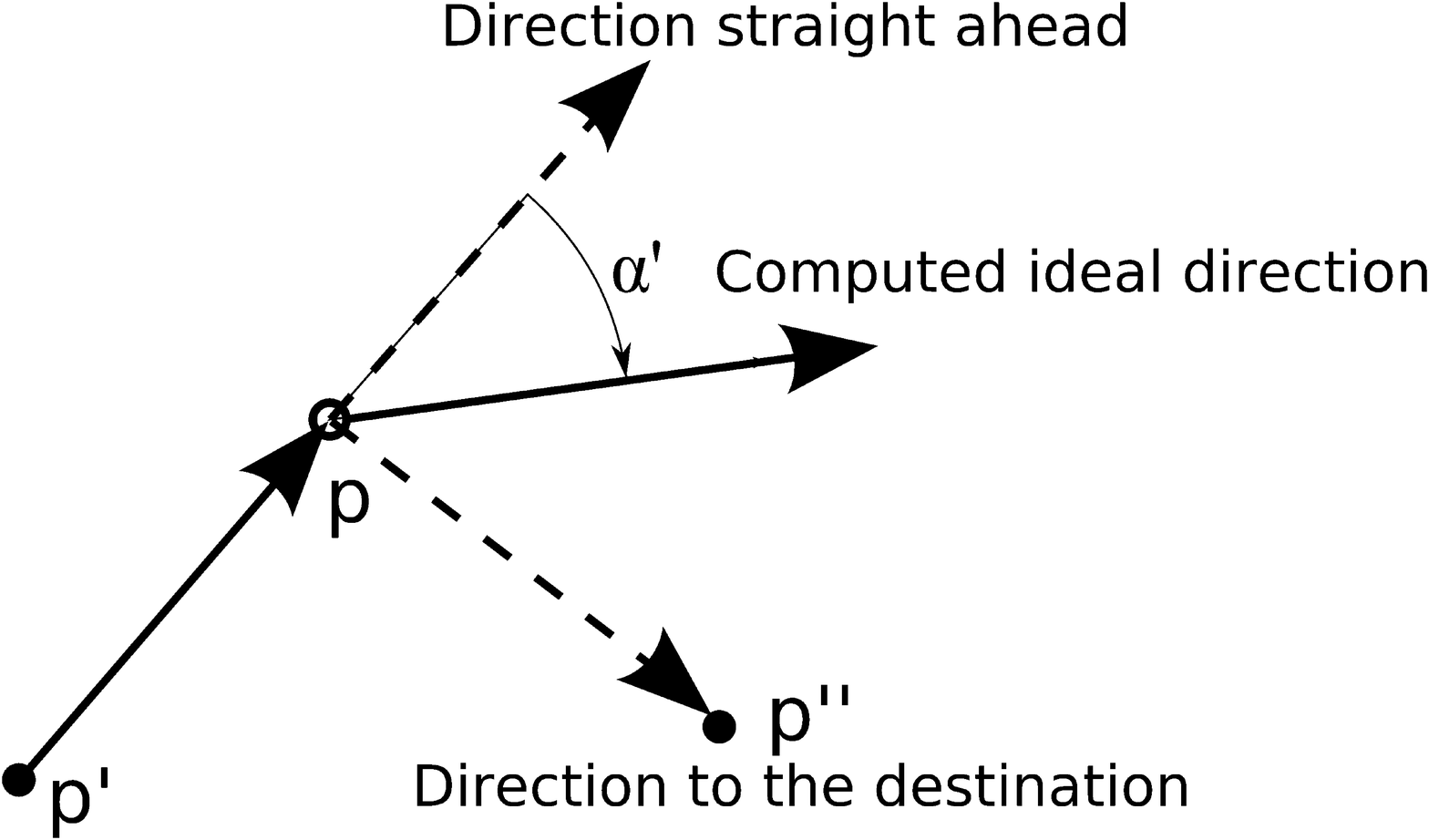}
}
\subfigure[The \texttt{compass} returns \texttt{NW}.]{
\label{fig compass2}
\includegraphics[angle=0,width=.35\textwidth]{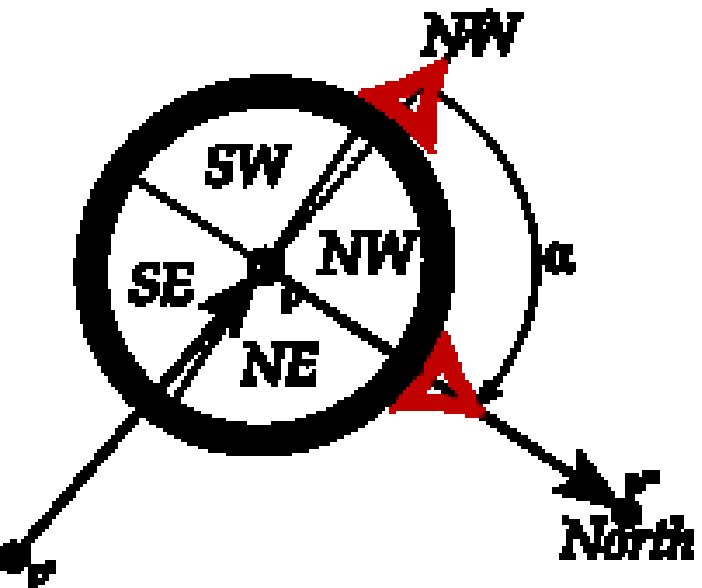}
}
\caption{Computation of the ideal direction and the \texttt{compass} device.}
\label{fig compass}
\end{figure*} 
The position $p'$ is assumed to be piggy-backed on the message,
therefore, access to $p$, $p'$ and $p''$ is consistent with the sensor
network model described in section \ref{sec: SN Model}.
First, the node $p$ computes $v_\textrm{prev} = \overrightarrow{p'p}$,  which is a vector
pointing in the last direction travelled by the message, as well as 
$v_\textrm{dest} = \overrightarrow{pp''}$, which is a vector pointing in the
direction of the message's destination. Through elementary
trigonometry an angle $\alpha$ is computed, such that $\alpha$ is the
unique angle in $\left[ -\pi, \pi \right [ $
with $\mathcal{R}_\alpha \cdot v_{\textrm{prev}}^\top = v_{\textrm{dest}}^\top$, 
where $v^\top$ denotes vector transposition and where
$\mathcal{R}_\alpha$ is the following rotation matrix:
\[
\mathcal{R}_\alpha =
\left(
\begin{array}{lr}
  \cos(\alpha) & -\sin(\alpha) \\
  \sin(\alpha) & \cos(\alpha)
\end{array}
\right) 
\]
The node can now compute the \emph{ideal direction}\footnote{
In the case where the node is the source of the message, $p'$ is not
defined so we simply set $v_{\textrm{ideal}} = v_{\textrm{dest}}$.} 
in which to
send the message. For the \emph{inertia mode}, the ideal direction is
defined as $v_{\textrm{ideal}} = \mathcal{R}_{\alpha'} \cdot v_{\textrm{prev}}$,
where $\mathcal{R}_{\alpha'}$ is a rotation matrix with $\alpha'$
defined by
\[
\begin{array}{lccl}
  \alpha' &=  - &\beta \pi& \textrm{ if } \alpha < - \beta\pi,\\
  \alpha' &=   &\beta \pi& \textrm{ if } \alpha >   \beta\pi,\\
  \alpha' &=  &\alpha&   \textrm{ otherwise}
\end{array}
\]
The ideal direction $v_{\textrm{ideal}}$ is thus obtained
by applying a rotation to the previous direction $v_{\textrm{prev}}$ towards
the destination's direction $v_{\textrm{dest}}$, however, the maximum
rotation angle allowed is bounded, in absolute value, by $\beta\pi$. 
This implies that the ideal direction is somewhere in between the
previous direction and the direction of the message's destination.
\begin{remark}
To enhance intuitive understanding of the \emph{inertia routing
mode}, it may be useful to notice that setting $\beta = 1$ implies
that $v_{\textrm{ideal}} = v_{\textrm{dest}}$. In other words the ideal
direction is always towards the message's destination and
\emph{inertia routing} is equal to the simple \emph{greedy geometric routing}.
At the other extreme, setting $\beta=0$ implies that
$v_{\textrm{ideal}} = v_{\textrm{prev}}$,
i.e. the message always tries to go straight ahead: there is
maximal inertia. In our simulations, setting $\beta=\frac{1}{6}$ proved to be a good choice
for practical purposes.
\end{remark}

After the ideal direction has been computed the message is sent to the
neighbour maximizing progress towards the ideal direction.
Maximal progress is defined to be reached by the node $m$ (amongst
outbound neighbours of the node currently holding the message) such that
$\textit{progress}(m):=
\left< v_{\textrm{ideal}}|\textit{pos}(m) - p\right>$ 
is maximized, where  $\left<\cdot | \cdot  \right>$ is the standard scalar product and $\textit{pos}(m)$ is
the position of $m$. 

\begin{figure*}[htb]
\centering
\subfigure[Success.]{
\label{fig inertia1}
\includegraphics[angle=0,width=.48\textwidth]{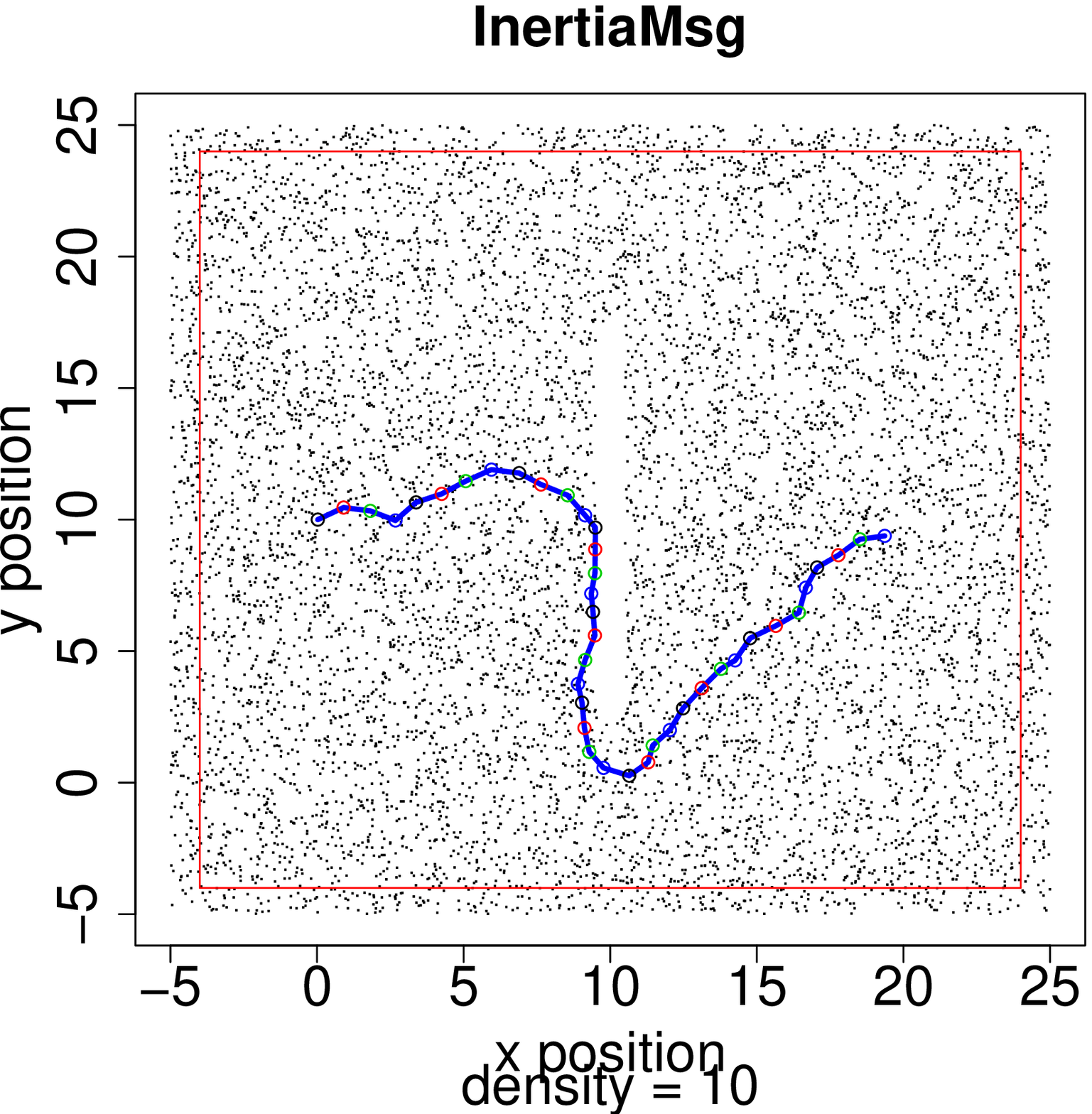}
}
\subfigure[Failure.]{
\label{fig inertia2}
\includegraphics[angle=0,width=.48\textwidth]{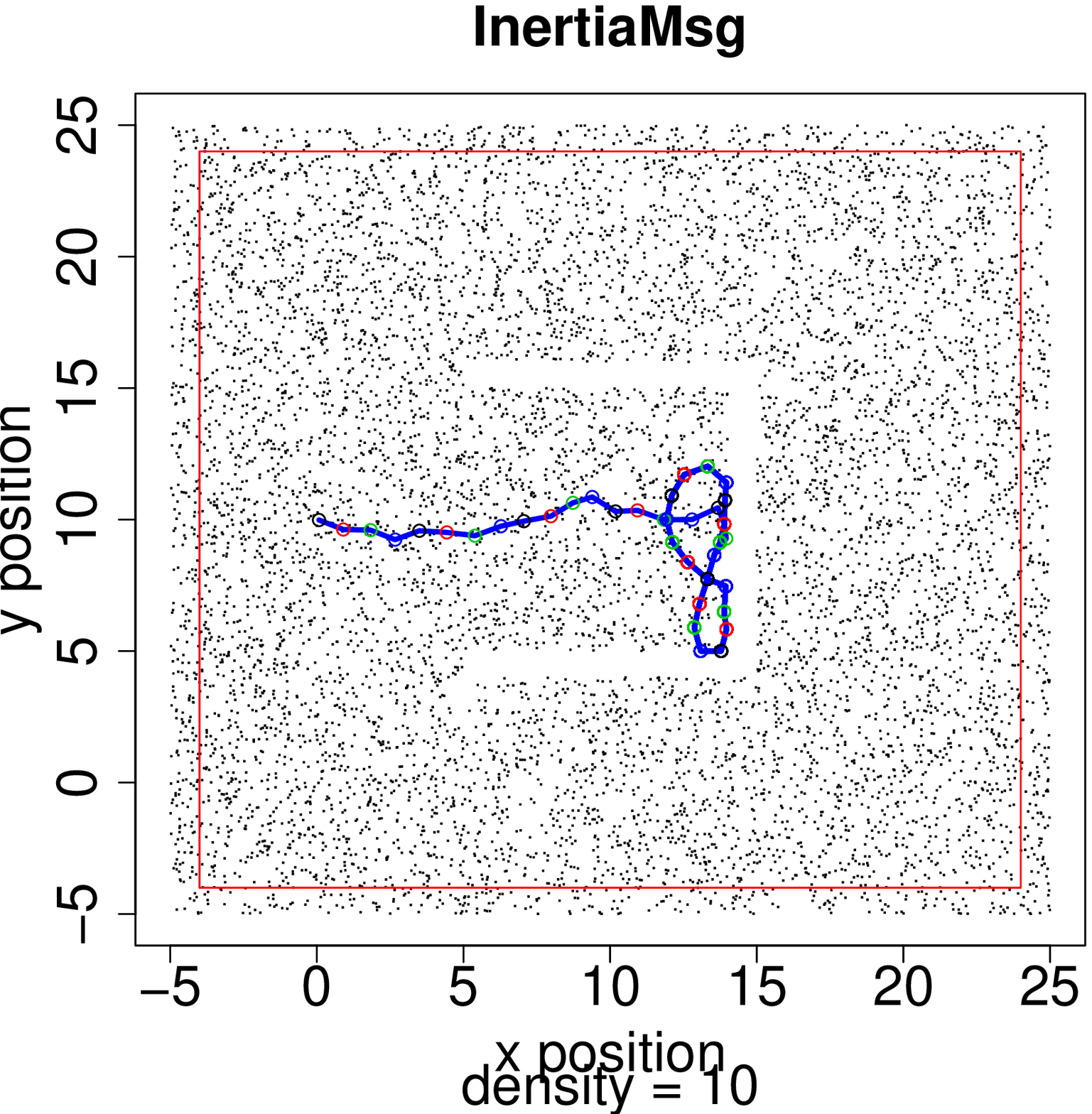}
}
\caption{Inertia routing with two different obstacles.}
\label{fig inertia}
\end{figure*}

Inertia routing is already quite an improvement over simple greedy
routing, as can be seen on figure \ref{fig inertia1} where a message
is successfully routed from a point $a=(0,10)$ to a point $b=(20,10)$
in the presence of an obstacle (a stripe). 
We delay detailed explanations of the simulation context in which the
plots of figure \ref{fig inertia} where obtained. Our intention in
showing figure \ref{fig inertia} now is to  show that although inertia
routing successfully avoids a simple obstacle
\footnote{We ask readers to
temporary believe us on the fact that the behaviour observed is
statistically representative; empirical evidence of this will be
given in section \ref{sec simulations}.}
(the stripe of figure \ref{fig inertia1}), it fails for the more complicated
U shape obstacle of figure \ref{fig inertia2} (more details will be
given in section \ref{sec simulations}). This routing
failure is the motivation for adding a \emph{contour mode} to which
our algorithm can switch when it needs to route messages around
complicated obstacles.

\subsection{Routing around obstacles:}
\label{route around obstacles}
We will now describe the \emph{contour mode} of our algorithm, as well
as the mechanism which permits to switch between \emph{contour mode}
and \emph{inertia mode}. The \emph{contour mode}, as well as the
\emph{switching mechanism} both make use of a \emph{virtual compass}
indicating the direction travelled by the message during its last hop
relatively to its final destination. 

\subsection{The compass device:}
\label{sec compass}
When a node generates a message, we take the convention to say
that the message has traveled its last hop towards the
north. Otherwise, we use the points $p$, $p'$ and $p''$ to compute the
angle $\alpha$ between the last direction travelled by the message
$v_{\textrm{prev}}$ and
the direction to the message's destination $v_{\textrm{dest}}$ as was
done in section \ref{sec inertia}, c.f. figure \ref{fig compass}.
Seeing $v_{\textrm{dest}}$ as the north, the virtual compass device
returns a value in north-west, north-east, south-west and south-east
depending on the quadrant in which $\alpha$ lies: 
\texttt{SW} if $\alpha \in \left[ -\pi, -\pi/2 \right[$,
\texttt{NW} if $\alpha \in \left[ -\pi/2, 0 \right[$,
\texttt{NE} if $\alpha \in \left[ 0, \pi/2 \right[$ and
\texttt{SE} otherwise.
See figure \ref{fig compass2} for
an intuitive explanation of the compass device, while a a rigorous
description of the computation steps implementing the virtual compass
is given in procedure \ref{algo compass} (\texttt{compass}).
\def\GETS{\leftarrow}
\begin{procedure}[hbt]
  \caption{Compass}
  \label{algo compass}
  \begin{algorithmic}
    \STATE $v_{\textrm{last}} \GETS \overrightarrow{p'p}$
    \STATE $v_{\textrm{north}} \GETS \overrightarrow{pp''}$
    \STATE $\alpha \GETS angle\_from\_to( v_{ \textrm{prev} }, v_{ \textrm{dest} } )$
    \IF{ $\alpha \in \left[ -\pi, -\pi/2 \right[$} 
	\RETURN \texttt{SW} 
    \ELSIF{ $\alpha \in \left[ -\pi/2, 0 \right[$} 
	\RETURN \texttt{NW} 
    \ELSIF{ $\alpha \in \left[ 0, \pi/2 \right[$}
	\RETURN \texttt{NE} 
    \ELSE[Comment: in this case $\alpha \in \left[ \pi/2, \pi \right[$] 
	\RETURN \texttt{SE}
    \ENDIF
  \end{algorithmic}
\end{procedure}
\begin{procedure}[hbt]
\caption{Mode Selector}
\label{algo switch}
\begin{algorithmic}
    \IF{ the flag is up}
      \IF{the flag is tagged with \texttt{E} \textbf{and} $\texttt{compass}\in\left\{ \texttt{NW}, \texttt{SW} \right)$}
        \RETURN contour mode
      \ELSIF{ the flag is tagged with \texttt{W} \textbf{and} $\texttt{compass}\in\left\{ \texttt{NE}, \texttt{SE} \right\}$) }
        \RETURN contour mode
      \ENDIF
    \ELSE
      \RETURN inertia mode
  \ENDIF
\end{algorithmic}
\end{procedure}

\subsection{The right and left hand rules to route around obstacles}
The inertia routing mode manages to route messages around some simple
obstacles, but not around more complicated obstacles. In order to
improve it, we need to detect when a message is trying to go around an
obstacle and switch to the \emph{contour mode} when appropriate. The
main idea in our contour mode is to choose either the \emph{right-hand
  rule} or the \emph{left-hand rule}, and to go around the obstacle
according to this rule. The right-hand rule means that the message
will go around the obstacle by keeping it on its right. When a node
receives a message, it makes a call to the \emph{compass} procedure to get the
direction of the message and uses a flag which is piggy-backed on the
message to decide if the right-hand rule (or the left-hand rule)
should be applied. If the flag is down, the algorithm looks at the
compass. If the compass points to the north, nothing special is done
(it means the message is making progress towards its destination, and
we stay in the inertia mode). But if the compass points south, it
means that the message is actually going away from its destination,
and the algorithm interprets this as the fact that the message is being
routed around an obstacle. To acknowledge this fact, \GRIC raises the
contour flag and tags it with \texttt{E} or \texttt{W} if the compass
points to the south-east or the south-west respectively. Once the flag
is up, the message is considered to be trying to go around an obstacle
using the right-hand rule or the left-hand rule if the flag is tagged
\texttt{W} or \texttt{E} respectively. See procedure  \ref{algo raise} (\texttt{raise flag})
for a detailed explanation. 
\begin{procedure}[hbt]
\caption{Raise flag (Raise and Mark the Flag)}
\label{algo raise}
\begin{algorithmic}
  \STATE\COMMENT{By assumption the flag is down}
  \IF{\texttt{Compass} == \texttt{SW}}
    \STATE Raise flag 
    \STATE Tag flag with the \texttt{W} value
  \ELSIF{\texttt{Compass} == \texttt{SE}}
    \STATE Raise flag
    \STATE Tag flag with the \texttt{E} value
  \ELSE
    \STATE Let the flag down
  \ENDIF
\end{algorithmic}
\end{procedure}
Once the message is considered to be
trying to go around an obstacle using the right-hand rule (i.e. if the
flag is up and tagged \texttt{W},), the algorithm will not change its
mind until the message reaches a node where the compass returns the
value \texttt{NW} (or \texttt{NE} if the left-hand rule was used). 
See procedure \ref{algo lower}
(\texttt{lower flag}) for a rigorous description. 
\begin{procedure}[hbt]
\caption{Lower Flag}
\label{algo lower}
\begin{algorithmic}
  \STATE\COMMENT{By assumption the flag is up}
  \IF{ ( flag is tagged with \texttt{W} \textbf{and} compass == \texttt{NW} )}
      \STATE put the flag down
  \ELSIF{ ( flag is tagged with \texttt{E} \textbf{and} compass == \texttt{NE} )}
      \STATE put the flag down
  \ELSE
    \STATE Let the flag up and don't change the tag
  \ENDIF
\end{algorithmic}
\end{procedure}
To summarize our discussion so far, when a node
receives a message two different cases may occur: either the flag is
down or it is up. If the flag is down (the message is not currently
trying to go around an obstacle), \GRIC runs procedure \ref{algo raise} (\texttt{raise flag}) 
to see if the message has just started going around an obstacle and
flag and tag it appropriately. On the other hand, if the flag is up,
it means the message was going around an obstacle, the procedure
\ref{algo lower} (\texttt{lower flag}) is used to see if the message
should be considered as having
finished going around the obstacle and the flag can be brought down.
The position of the flag, together with the position of the compass
is used to determine if the message is to be routed according to
the \emph{inertia} or the \emph{contour} mode.
\subsection{Mode selection}
\label{sec switch}
The flagging mechanism described in the previous section enables a
node to know if a message is being routed
around an obstacle by looking at the position of the flag (which is
piggy-backed on the message). 
When the flag is down, the \emph{inertia mode}
is used. However, when the flag is up, the \emph{contour mode} (which we shall
describe below) \emph{may} have to be used. We next
describe informally the mechanism used to decide whether the contour mode or
normal mode should be used \emph{when the flag is up}. 
Therefore, suppose the flag is
up, and in order to simplify the discussion, suppose it is tagged with
the \texttt{E} value (if the tag is \texttt{W}, a symmetric case is
applied). Note that by case assumption, this implies that the compass
either returned \texttt{NW}, \texttt{SW} or \texttt{SE}, since
otherwise procedure \ref{algo lower} (\texttt{lower flag})
would have put the flag down. The message is therefore currently
trying to go around an obstacle using the left-hand rule described in
the previous section,
trying to keep the obstacle on its left. Two different
cases have to be considered. \underline{In the first case}, the compass points
\texttt{SE}. Recalling the definition of the \emph{ideal direction}
$v_{\textrm{ideal}}$ and of the previous direction $v_{\textrm{prev}}$ 
from section \ref{sec inertia}, it is easy to see that since by case
assumption the compass points to south-east,  $v_{\textrm{ideal}}$ is
obtained by applying to $v_{\textrm{prev}}$ a rotation to the
left 
(i.e. counter-clockwise),
which is equivalent to saying the angle $\alpha$ defined in
section \ref{sec inertia} takes a value in $\left[0,\pi\right]$
and therefore that the angle $\alpha'$ takes a value in
$\left[0,\pi/6\right]$). 
In other
words, the \emph{inertia routing mode} tries to make the message
turn left, which is consistent with the idea of the left-hand
rule: routing the message along the perimeter of the obstacle
while keeping it to the left of the message's path. There is therefore
no need to switch to the rescue \emph{contour mode}. \underline{The
second case} which may occur is when the compass points
either to the \texttt{SW} or to the \texttt{NW}. A similar reasoning
to the previous one show that in this case
the inertia routing mode will make the message turn right and the
angle $\alpha'$ will be in
$\alpha'\in\left[-\pi/6,0\right]$. According to the left-hand rule
idea, the obstacle should be kept on the left of the message, however
turning right, according to the inertia routing mode,
will infringe the idea of keeping the obstacle to the left. Instead,
the message would turn right and get away from the perimeter of the
obstacle. In this case, it is therefore required to call the rescue
mode: the \emph{contour mode}.
Rigorous description of the above discussion is given in
procedure \ref{algo switch} (\texttt{mode selector}).
\subsection{The contour mode}
By case assumption, the flag is up.
We also suppose without loss of generality that the tag on the flag is 
\texttt{E}, since the case of a \texttt{W} tag is similar.
We have seen that procedure
\ref{algo switch} (\texttt{mode selector}) calls the contour mode when the inertia mode is
going to make the message \emph{turn right} (or \emph{left} if the tag is
\texttt{W}),
which is equivalent to saying that
$\alpha\in\left[-\pi,0\right]$ and $\alpha'\in\left[-\pi/6,0\right]$.
However, the left-hand rule would advice turning on the left to stay as
  close as possible to the obstacle and to keep it on the left of the
  message path.
Therefore, in order to stay consistent with the left-hand rule idea, we define
$\alpha_2 = -\textit{sign}(\alpha)(2\pi - \left|\alpha\right| ) $,
  thus $\alpha_2$
is
an angle such that $\textit{sign}(\alpha_2) = -\alpha$ and
such such that $R_{\alpha_2}\cdot v_{\textrm{prev}}^\top = R_{\alpha}\cdot
v_{\textrm{prev}}^\top = v_{\textrm{prev}}^\top$.
In order to give some inertia to the message we define $\alpha_2'$
from $\alpha_2$ in a similar way as $\alpha'$ was defined from
$\alpha$, 
and the ideal direction $v_{\textrm{ideal}}$, is defined by 
$v_{\textrm{ideal}}^\top = R_{\alpha_2'}\cdot
 v_{\textrm{prev}}^\top$, where $\alpha_2' = \beta\alpha_2$, where $\beta$
 is the inertia conservation parameter of section \ref{sec  inertia}. 
Putting all things together, \GRIC is formally described by the
(non-randomized version of) algorithm
\ref{algo GRIC}.
\begin{algorithm}
\caption{\GRIC, running on the node $n$ which is at position $p_n$.}
\label{algo GRIC}
\begin{algorithmic}[1]
  \IF{the flag is down}
    \STATE call \texttt{\bf{Raise flag}} \COMMENT{Procedure \ref{algo raise}}
  \ELSE
    \STATE call \texttt{\bf{Lower flag}} \COMMENT{Procedure \ref{algo lower}}
  \ENDIF
  \STATE $mode := \texttt{\bf{Mode selector}}$ \COMMENT{Procedure \ref{algo switch} }
  \IF{ $mode == \textit{inertia mode} $ }
    \STATE $\gamma := \alpha'$ \COMMENT{Where $\alpha'$ is the angle defined in section \ref{sec inertia}} 
  \ELSE
    \STATE $\gamma := \alpha_2'$ \COMMENT{Where $\alpha_2'$ is the angle defined in section \ref{route around obstacles}} 
  \ENDIF
  \STATE $v_{\textrm{ideal}}^\top := R_\gamma\cdot v_{\textrm{prev}}^\top$
  \STATE \label{random line}Let $\mathcal{V}$ be the set of neighbours of $n$.
  \IF{running the non-random version of \GRIC}
    \STATE Send the message to the node $m\in\mathcal{V}$ maximizing
    $\left< v_{\textrm{ideal}}|\overrightarrow{p_n p_m}\right>$, where
    $p_m$ is the position of $m$.
  \ELSIF{running the random version of \GRIC}
    \STATE Let $\mathcal{V'}$ be an empty set.
    \FORALL{ $v\in\mathcal{V}$ }
      \STATE add $v$ to $\mathcal{V'}$ with probability $0.95$
    \ENDFOR
    \STATE Send the message to the node $m\in\mathcal{V'}$ maximizing
    $\left< v_{\textrm{ideal}}|\overrightarrow{p_n p_m}\right>$, where
    $p_m$ is the position of $m$.
  \ENDIF
\end{algorithmic}
\end{algorithm}

\subsection{Randomization}
\label{randomization}
We shall show experimentally
that \GRIC succeeds in routing messages around local minimum and around
obstacles with high probability (where the probability is taken over
randomly generated networks). However, in unfavorable cases it may fail. We
observed experimentally that adding a small random disturbance to the
\GRIC algorithm improves its behavior. Intuitively, if the message
gets blocked in a local minimum and starts looping, a small random
perturbation may be sufficient to make the message escape the routing
loop.  
Instead of considering $\mathcal{V}$ to be the set of
neighbours of $n$, we consider $\mathcal{V}$ to be the \emph{subset}
of neighbours of $n$ where each neighbour is added to $\mathcal{V}$
with probability $1-\epsilon$, where $\epsilon$ is a small constant. For practical purposes experiments have
shown the choice of $\epsilon = 0.05$ to be good. 
Formally, this idea is implemented by introducing the if-else cases on
lines 12 and 14 of algorithm \ref{algo GRIC} to
differentiate between random and non-random routing modes.
It may be worth noticing that the random version of \texttt{GRIC}
could also be interpreted as simulating temporary link failure in the
network. This point of view implies that \texttt{GRIC} is actually
robust to limited link failure.
\chapter{Experiments}
\section{Methodology}
\label{sec simulations}
Our methodology is the following.
We abstractly design a simple experiment which consists of (1)
deploying a sensor net, (2) generate a message which has to be routed
from a point $a$ to a point $b$ of the network and (3) measure the
outcome of the experiment (successful or failed message delivery and efficiency of
the routing path found by our algorithm, but c.f. details below). Running
experiments on a real sensor network has a high cost (sensor
nodes are still prohibitively expensive to run large scale
experiments) and even if using a small scale real sensor network, running
enough experiments to have statistically meaningful results
would be extremely time consuming and probably unrealistic in the
design phase of the algorithm. For those reasons, we decide to use ``in silico'' simulations to
validate our sensor network algorithm which we describe below.
A single simulation consists of deploying sensor nodes randomly and
uniformly in a square region $\mathcal{R}$ of $30\times 30$ unit length
sides and routing a message from
the point $a=(10,0)$ to the point $b=(20,0)$ (c.f. figures \ref{path
  stripe} to \ref{path cuve bad}).
\subsection{Simulation Platform and Model}
For computational efficiency of our simulation platform, we
use a high level simulation of a sensor network. In particular, we do
not simulate the physical, the MAC and the network layer of the sensor
network, but use the unit disc graph combinatorial model of a real
world communication graph. In this model, each sensor is a point in
the plane and the wireless links are between two sensor which are at
at most at distance 1 from one another. The unit disc graph model is
widely used and is a reasonable mathematical abstraction of a sensor
net \emph{for simulation purposes only}. That is, we would like to
emphasise that although we use the unit disc graph for our
simulations, we think that the behavior of our algorithm would be
significantly similar in the case of a real world network 
where
wireless links are supposed to be made
available by the link layer protocol (c.f. section \ref{sec: SN Model}
for a detailed description) 
or in the
case of simulations conducted with a more realistic (but
computationally more expensive) communication graph model such as the
radio irregularity (RIM) model from \cite{ZHK04} or the signal to
noise ratio(SNR) model.  The parameters of an experiment are
(1) the routing algorithm used, (2) the
obstacles added to the network, (3) the density of the network.
We compare the FACE algorithm, the greedy algorithm, the inertia
algorithm, the LTP algorithm and the \GRIC algorithm (both the random and the non random
version)
\label{algorithmsTested}
\section{Algorithms tested}
We test the \GRIC algorithm. 
When we want to explicitly state whether
we are talking of the randomized or the non randomized version, we use
\GRICplus for the random algorithm and \GRICminus for the non-random. 
\GRICplus outperforms the \GRICminus.
However, because we understand that when
\GRICminus successfully routes a message to the
destination it does it using a short path, we use \GRICminus as a
lower bound for the hop count. We would use greedy, however greedy
fails to route around obstacles. (Actually, greedy is used when there
are no obstacles, c.f. figure \ref{metric void} and \ref{metric
void2}.
On the other hand, \GRICplus may take more time to route messages to
their destination than \GRICminus does because its randomized component could, in bad
cases, make it loop for a long time inside a blocking obstacle before it finally,
by chance, ``jumps'' out of the obstacle following a favorable random
outcome.
 
We also
consider the FACE algorithm, which is used as an upper bound on the
success rate (since it
guarantees delivery for all possible obstacle shapes as long as the
communication graph is connected). The inertia algorithm from section
\ref{sec inertia} is also being simulated. Inertia is similar to
\GRIC, except that it does not use the right-hand rule. It is
therefore a simplified version of the \GRICminus algorithm. As a
consequence, it fails except for the simplest obstacle (the stripe
shape obstacle of figure \ref{obstacles}). We also consider the greedy
algorithm, as a lower bound on the success rate (greedy is
known to fail in the presence of routing holes) and a lower bound on
the hop count and travelled distance since, greedy being greedy, it
routes very quickly to the destination. 

For comparison purposes with
previous work, we also include the LTP algorithm from \cite{CNS02,CNS05}
which combines a randomized greedy forwarding with backtracking in
the case where messages get blocked in local minimum. (In particular,
comparing to LTP permits to compare to other randomized algorithm
which were tested against LTP in \cite{CMN06}). \underline{LTP is a good choice
for comparison purposes} because
\cite{CMN06} finds that it outperforms the other tested protocols in
conditions similar to the ones we consider. (The VTRP protocol from
\cite{ACM04} actually competes in the case where no obstacles are
present, but it does so by increasing the transmission range of the nodes).
The obstacles we consider are either no obstacle at all, the stripe,
the U shaped and both concave shapes (concave shape 1 and concave shape
2) which can be seen on figure \ref{obstacles} (the case with no obstacle is
not plotted on this figure). We consider network
densities ranging from \emph{low} to \emph{high} (c.f. figure
\ref{densities}), and section \ref{density discussion} of the appendix
for a
discussion on network density.
In order to
be statistically meaningful, for each routing algorithm considered,
for each density considered and for each type of obstacle considered
we repeat $1000$ times the experiment described above (dispersion of
$n$ nodes, computation of the communication graph based on the unit
disc mode and routing of a single message from $a$ to $b$ until either
the algorithm succeed or fails). We consider the following metrics to
measure the quality of an algorithm: the \emph{success rate}, which is
the success to failure ratio of the algorithm (success meaning that
the message reaches its destination), the median \emph{hop
  distance} as well as the median \emph{Euclidean distance} travelled
by messages on successful routing outcomes. The hop count is a common measure
of quality for routing algorithms because of its broad interpretation range:
it measures the number of messages 
exchanged and thus the \emph{in-network traffic}, it is also a simplified
measure of \emph{energy consumption} (at least if the transmission power is
fixed and does not depend on the distance of the transmission), as
well as a measure of \emph{time} or \emph{latency}, in the sense that we can assume
that data transmission takes a close to constant amount of time (at
least for a fixed message length) and we use it to verify that
routing is not only successful but also efficiently.
\section{Digression on the network density}
\label{density discussion}
The \emph{network density} is, after the choice of the routing algorithm to
be tested, the most important network parameter of our simulations.
It is therefore natural to choose carefully what network densities are
going to be considered for our simulations. 
The \emph{network density} is defined as the average number of sensors per unit
square area. It is well known to be an important network parameter for
georouting algorithm with a strong influence on success rates (except
for the FACE family of algorithms, which guarantee delivery for all
densities as long as the network is connected).
As confirmed by our simulations, high network density tends to
increases success rates (i.e. the
probability that a message reaches its destination) for all the
algorithms we considered.

Since high density improves success rate, it is desirable to have
dense networks. This can be achieved by augmenting the number
of sensor nodes (which thus implies an extra cost, since more devices
have to be used), alternatively the transmission power of sensor
nodes can be increased (since increasing the transmission power will
increase the transmission range and thus the connectivity of the
communication graph). This alternative solution also has a cost in
terms of energy since increasing the transmission power will mean that
sensor nodes will deplete their energy faster and thus the network
lifetime will be diminished. It is therefore evident that although
high density is desirable for georouting algorithm to successfully
deliver messages to their destination, the constraints imposed by
sensor nets imply that high densities come at an extra energy or monetary cost. 
An important quality measure is therefore appearing: high success rate
should be achieved even for low densities.

This discussion makes it clear that it is important to understand what
are reasonable assumptions about the network density. Our simulations
are made using the unit disc graph model. This is a common and
reasonable combinatorial abstraction of a real world communication
graph for simulation purposes. In order to compare with other models
and other work (typically using a non-unit disc graph model, i.e. when the
transmission radius is not $1$ but some other constant value $R$), it
is useful to recall that the density $d$ is directly related to the
average number of neighbours in the communication graph:
the average number of neighbours in the unit disc graph
is approximately $\approx \pi\cdot d$. We now turn to examine what are
reasonable densities for sensor network simulations using the unit
disc graph.

Sensor networks are usually
considered to be dense networks, but this informal assertion is rarely
made precise.
We argue that, although being somehow arbitrary, the following
interpretation of densities is reasonable: densities below $3$
should be considered \emph{very low} densities, while $d=3$ is a 
\emph{low} density, $d=4.5$ is a \emph{medium} density, $d=6$ is a \emph{high density} and
densities above $8$ are \emph{very high}. 
Those convention are based on observing the
network densities used in related work as well as on observing
experimentally the connection probability of the graph.
While presenting the FACE algorithm, \cite{SD04} experiments with an average number of neighbours between $4$
and $10$ (i.e. densities between $\approx 1.3$ and $\approx
3.2$). Those are very low densities, since our simulations show
(c.f. left-most plot in figure \ref{metric void})  that
the success rate of the guaranteed delivery FACE algorithm drops when
the density is below $3$.
Since FACE has guaranteed delivery, failure follows from the
communication graph being disconnected\footnote{Actually, in our
  simulation setting this could also follow from the message going
  \emph{out of bounds}, but see section \ref{simdetails} for the exact definition of
failure in our experiment.}.
The phase transition phenomena we observe is consistent with the \emph{caveat to
  connectivity} discussion from 
\cite{KW05c} 
and the findings of \cite{KWB03}.
 
The reason \cite{SD04} manages to work with such low densities is that
they generate random graph but remove disconnected instances
(i.e. most of the random outcomes) from their
experiments. This is an artificial procedure in typical scenarios for sensor
nets, where nodes are considered to be deployed somehow unattended
(possibly even dropped from a plane), but it could be reasonable 
for nets where the node deployment is managed and guarantees connectivity.
Because the
success probability of FACE (as well as for \GRIC~) is close to $1$ for
densities of $3$, we
consider densities of $3$ to be \emph{low} but reasonable. Because
even the guaranteed delivery FACE algorithm rapidly starts failing 
when densities drop below $3$, we consider them to be \emph{very low}.
We consider densities of $6$ to be \emph{high} because it is twice the low density. 
Although being high, densities of $6$ are still reasonable as
acknowledged by their use in some of the relevant and recognised
related work (c.f. details below).
Having settled for the low and high densities to be $3$ and $6$
respectively, we consider densities in the middle, i.e. densities
around $4.5$, to be medium densities. Finally, we consider densities
around $8$ to be \emph{very high}. Our discussion is summarized on
table \ref{densities}. We feel that this interpretation of network
densities is fairly conservative (i.e. we do not allow ourselves to
work with too high densities), and we run our simulations accordingly,
letting network density range from very low to very high ($d=0$ to $d=10$). 
\\
\subsection{ Some densities used in seminal and relevant previous work}
In the
celebrated direct diffusion paper \cite{IGE03}, a low density of $3.125$
is used for experiments.   
In their proposal of georouting with virtual coordinates, \cite{RRP03}
experiments are conducted with an average number of neighbours equal to
$15$ and thus a medium density of $\approx 4.8$. In their study of the
impact of obstacles on different routing algorithms \cite{CMN06}, after discussing
the matter, the authors settle for a high density of $6.25$.
A high density is also used in \cite{KNS06}, where most simulation are
conducted with an average number of neighbours equal to
$20$, which is equivalent to a density of $\approx 6.3$ in our
setting.
\begin{figure*}
\centering
\begin{tabular}{|l|c|c|c|c|c|}
\hline
\textbf{Density interpretation} &  very low  &  low & medium & high
density & very high density \\
\textbf{Density value} & $< 3$ & $\simeq 3$ & $\simeq 4.5$ & $\simeq 6$ & $\geq 8$ \\
\hline
\end{tabular}
\caption{Typical densities and their interpretation in the context of
  sensor nets.}
\label{densities}
\end{figure*}
\section{Simulation details}
\label{simdetails}
\subsection{Simple experiment}
A \emph{simple experiment} consists in randomly deploying sensors in a square
region $\mathcal{R}$ and to generate a single message which has to be
routed from a point $a$ to a point $b$. An experiment is defined by a
single parameter $d$ which is the network density, and it may have two
possible outcomes: success or failure. We next describe in more
details but with some level of abstraction our experiment design. 
The square region $\mathcal{R}$ is chosen to be a square region of side $30$ defined, in
$x$-$y$ coordinates, as the set of points $\mathcal{R}~=~
\left\{(x,y)\in\mathbb{R}^2|-5\leq x \leq 25\textrm{ and }-5\leq y \leq
25\right\}$. Given the density $d$, we randomly and uniformly deploy $n=d\cdot
900$ sensor nodes in the region $\mathcal{R}$. Each node is assumed to
know its location as well as the list of its neighbours and their
location (c.f. details below below). 
A simple experiment is also defined by the routing algorithm we want
to test. Each node runs the same algorithm (for example the \GRIC,
the greedy or the FACE algorithm). Once this is done, a single message
is generated at the point $a=(0,10)$. Since there may not be any node
exactly at this point, the message should be attached to the node  which is
the closest to the point $a$. The message has a destination, which is
the point $b$ with coordinates $b=(20,10)$. Network data propagation is
simulated by letting the message being propagated from node to node, the
next hop being chosen by the routing algorithm running on the
nodes. The experiment goes on until it is either decided that the
routing was successfully or that it has failed. Routing is considered
\emph{successful} if the message is sent to any node which is at
distance less than $1$ from the destination point $b$. There are two
conditions for terminating the simulation with a conclusion that the
algorithm has failed. The first one is if the message has not reached
its destination after a given time to live TTL. The TTL is fixed at a
very high value (TTL$=n$, where $n$ is the number of nodes). Because
the TTL is very high, we could have some successful outcomes where the
routing is very inefficient (i.e. the message does indeed reach its
destination, but after a very long time). To be sure that this is not
the case, we also keep track of the distance travelled by the message
(both the Euclidean distance and the hop count), and verify that our
algorithm not only routes messages to there destination, but that it
those so using a short path. The other condition under which we
consider the routing to be a failure is if the message gets within
distance $1$ of the border of $\mathcal{R}$. This second condition for
failure is a precaution, to ensure that routing protocols do not use
the border of the region $\mathcal{R}$ to successfully route messages
to their destination.

\subsection{ Summary of simulation results}
\begin{figure*}[htb]
\centering
\subfigure[Success, concave shape2.]{
  \label{metric cuve bad}
  \includegraphics[angle=0,width=.48\textwidth]{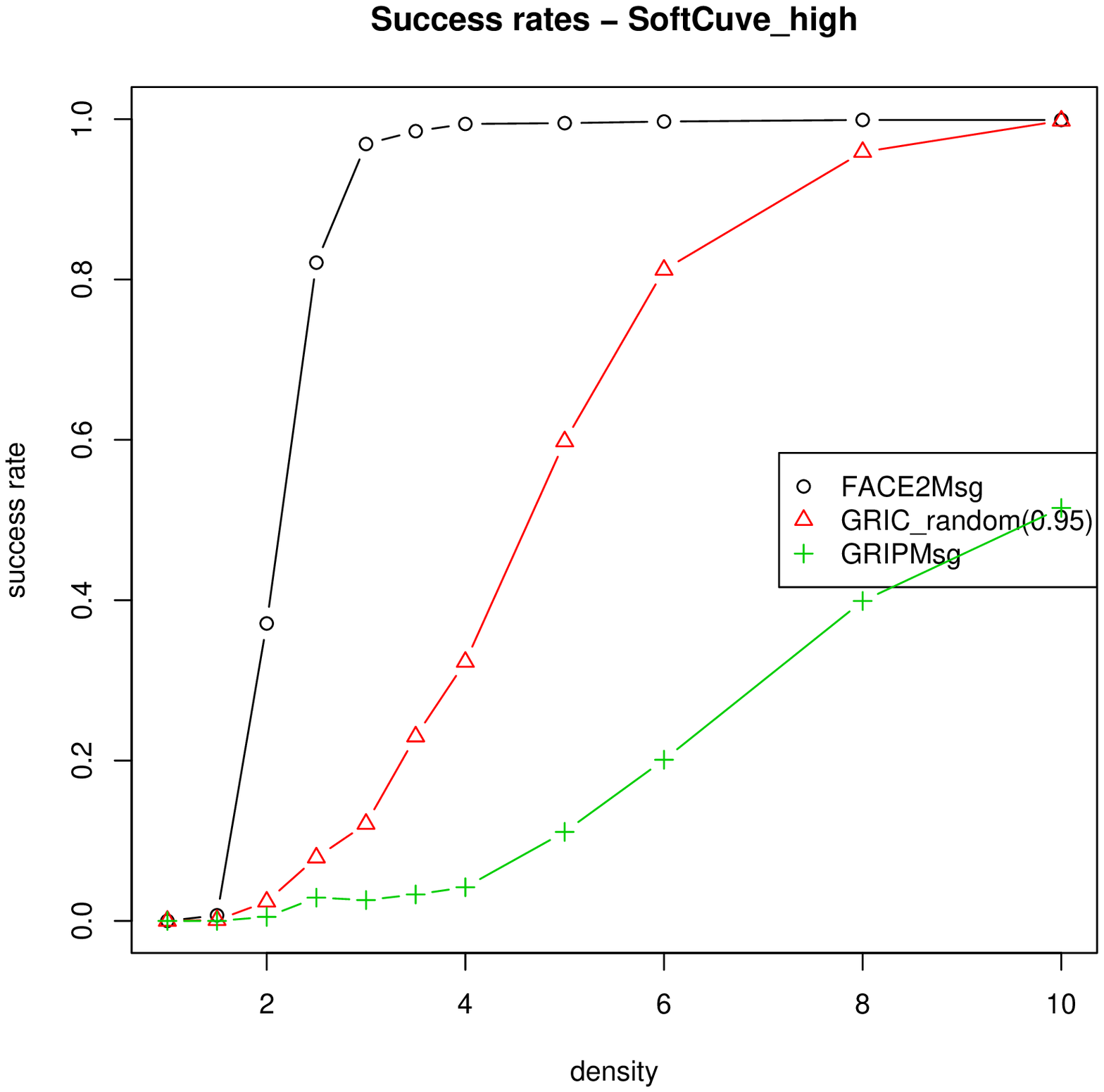}
}
\subfigure[Hops, concave shape2.]{
  \label{metric cuve bad2}
  \includegraphics[angle=0,width=.48\textwidth]{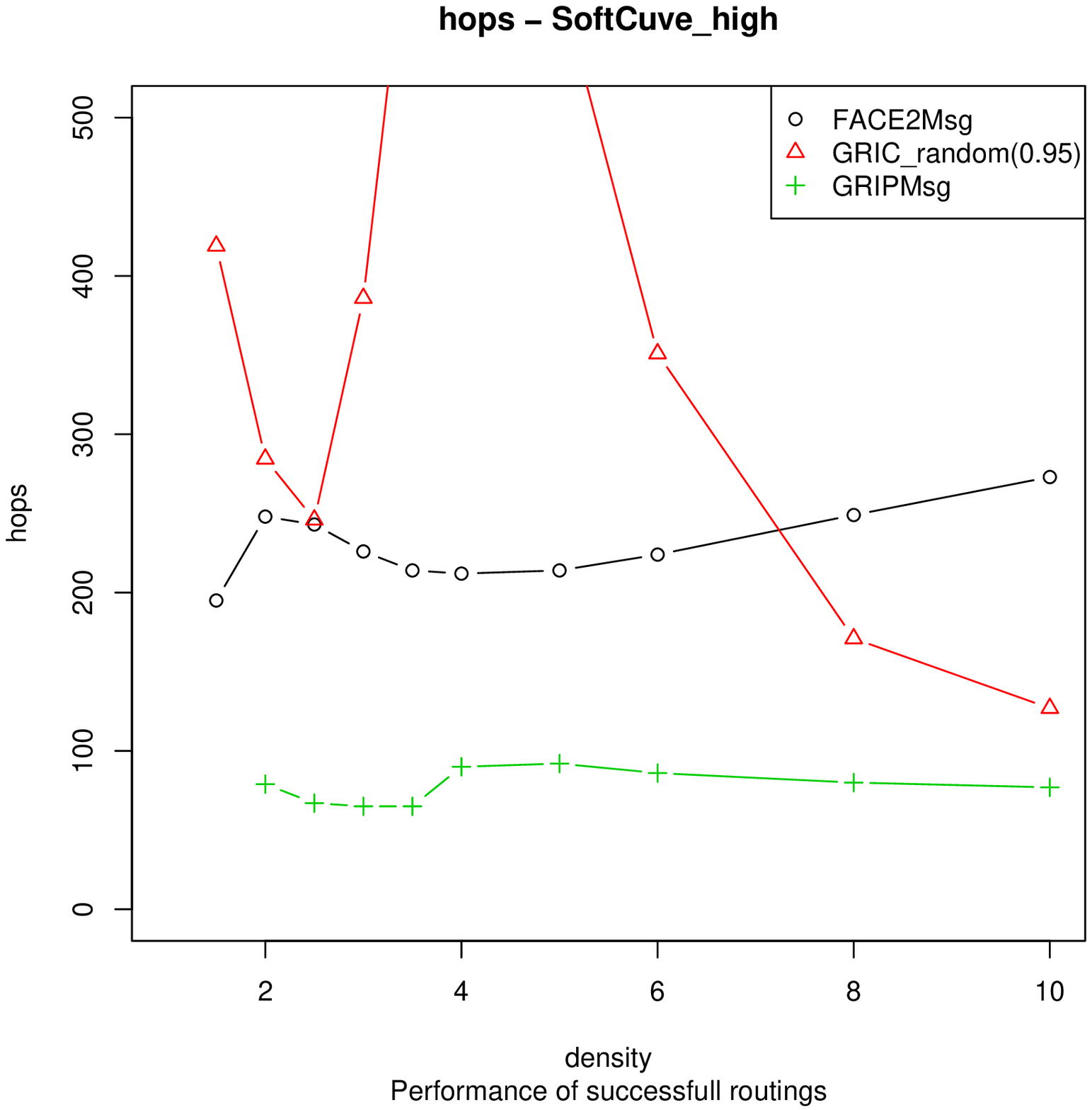}
}
\caption{Success rate and hop-count when varying the network density.}
\label{fig simulation results}
\end{figure*}

We first look at the case where no obstacles are added\footnote{
Figure \ref{fig simulation results} summarizes experimental results.
LTP5Msg stands for the LTP algorithm and GRIPMsg stands for the non-randomized version
of \GRIC, which is given essentially for comparison purposes and to
give a lower bound on the hop count. Further experimental results are given in section
\ref{appendix figures} of the appendix.}
.
Figure \ref{metric void} shows the impressive success rate of \GRIC: it
delivers messages with almost 100\% success rate on the condition
that the network is connected. This assertion follows from the fact that
FACE is analytically proved to have 100\% success as long as the
network is connected, and \GRIC performs almost as well as
FACE\footnote{The attentive reader may have noticed that for
the extremely low density of $d=1$ \GRIC seems even better than
FACE. This is possible in our experiment setting due to the ``out of
bound'' mechanism that stops messages being routed on the perimeter of
the network, c.f. previous section for details.}.
Looking at the path length (figure
\ref{metric void2}), we see that \GRIC performs very well (as well as
greedy). Because \GRIC requires no planarization, it seems to combine
the ``best of two worlds'': the simplicity (and short paths) of greedy
and the success rate of FACE. These results indicate
that in the absence of obstacles, \GRIC outperforms other georouting routing algorithm.
 
\begin{figure*}[htp]
\centering
\subfigure[Stripe.]{
  \label{path stripe}
  \includegraphics[angle=0,width=.48\textwidth]{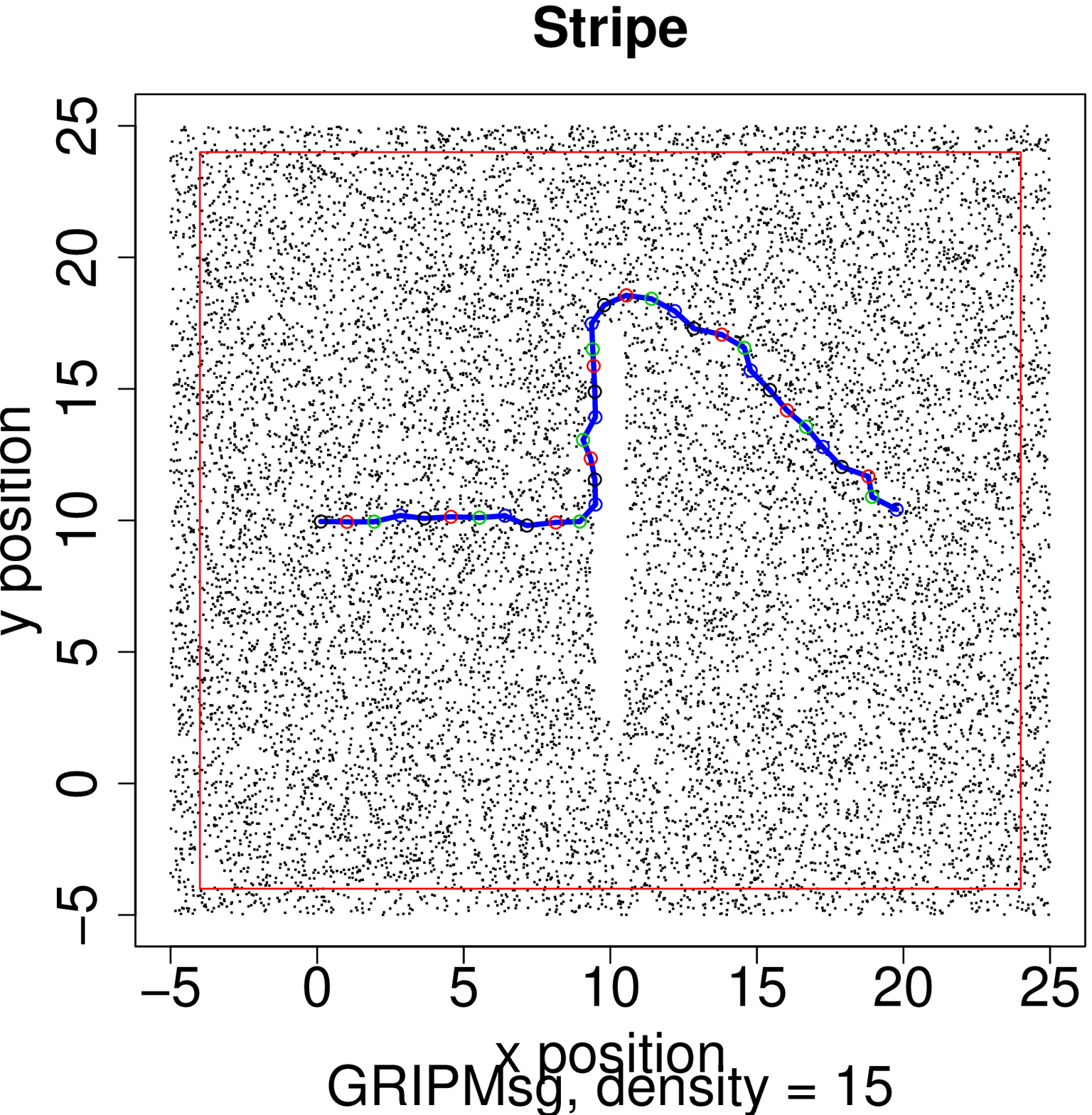}
}
\subfigure[U shape.]{
  \label{path U}
  \includegraphics[angle=0,width=.48\textwidth]{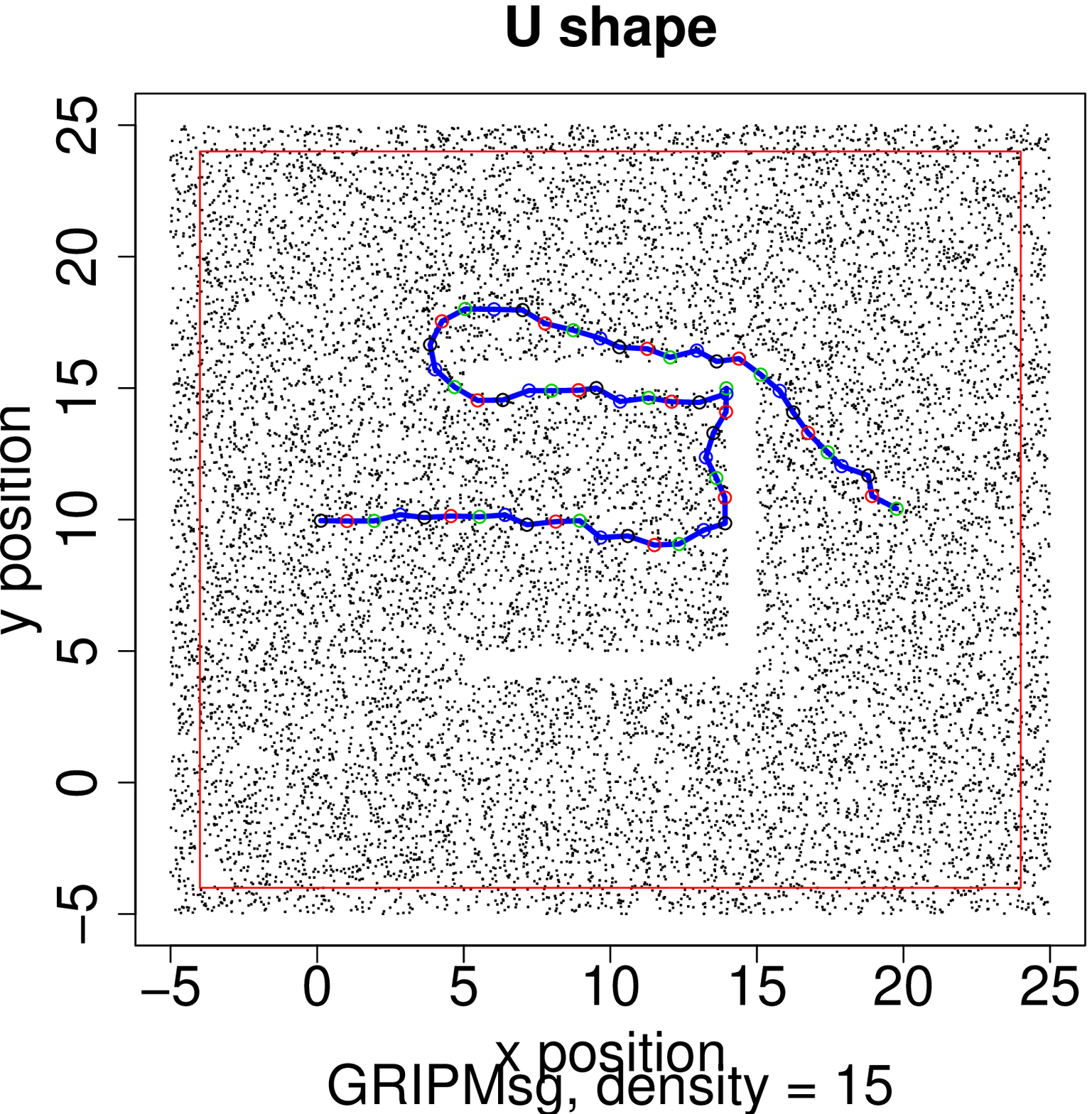}
}
\subfigure[Concave shape 1.]{
  \label{path cuve good}
  \includegraphics[angle=0,width=.48\textwidth]{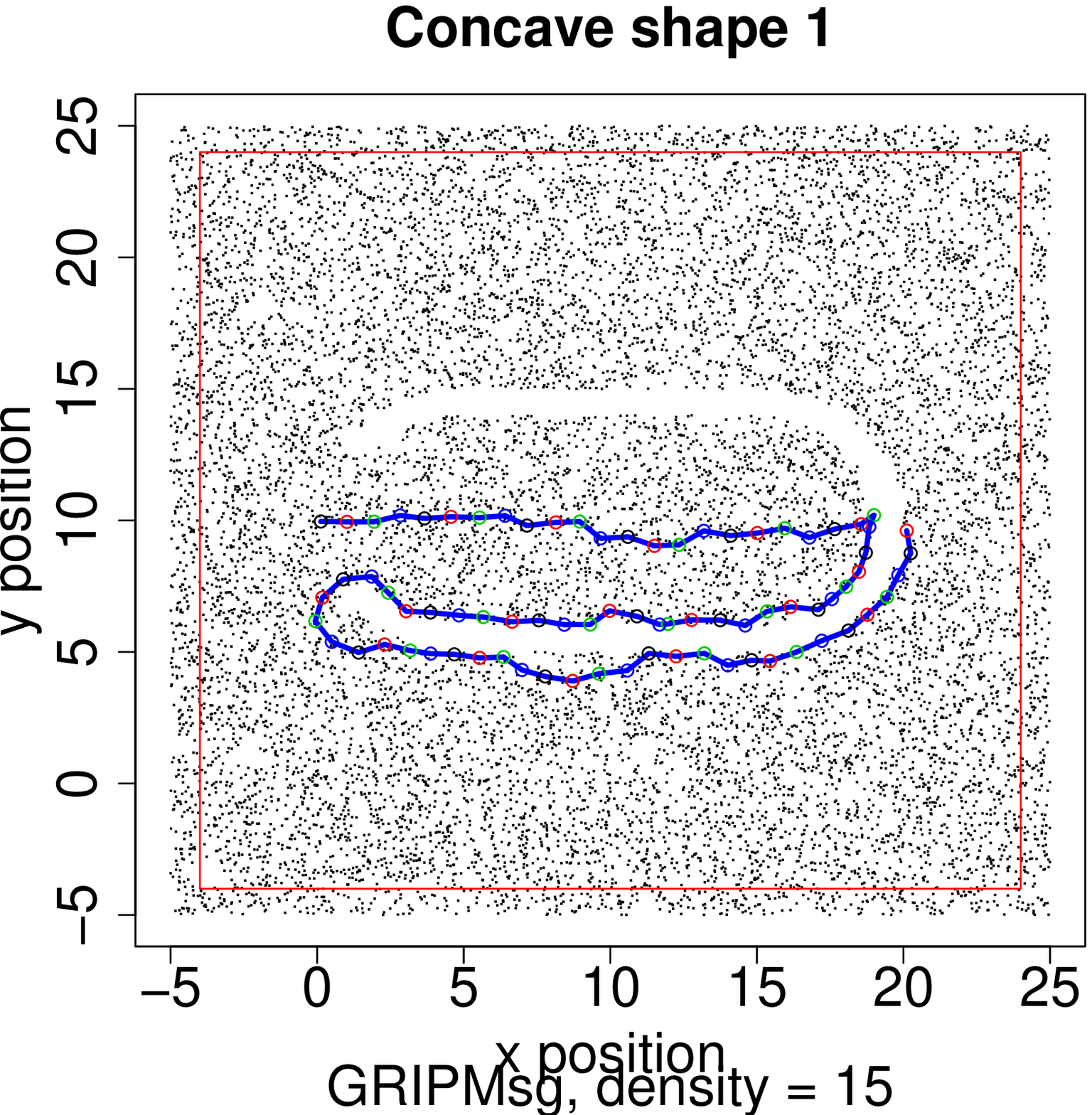}
}
\subfigure[Concave shape 2.]{
  \label{path cuve bad}
  \includegraphics[angle=0,width=.48\textwidth]{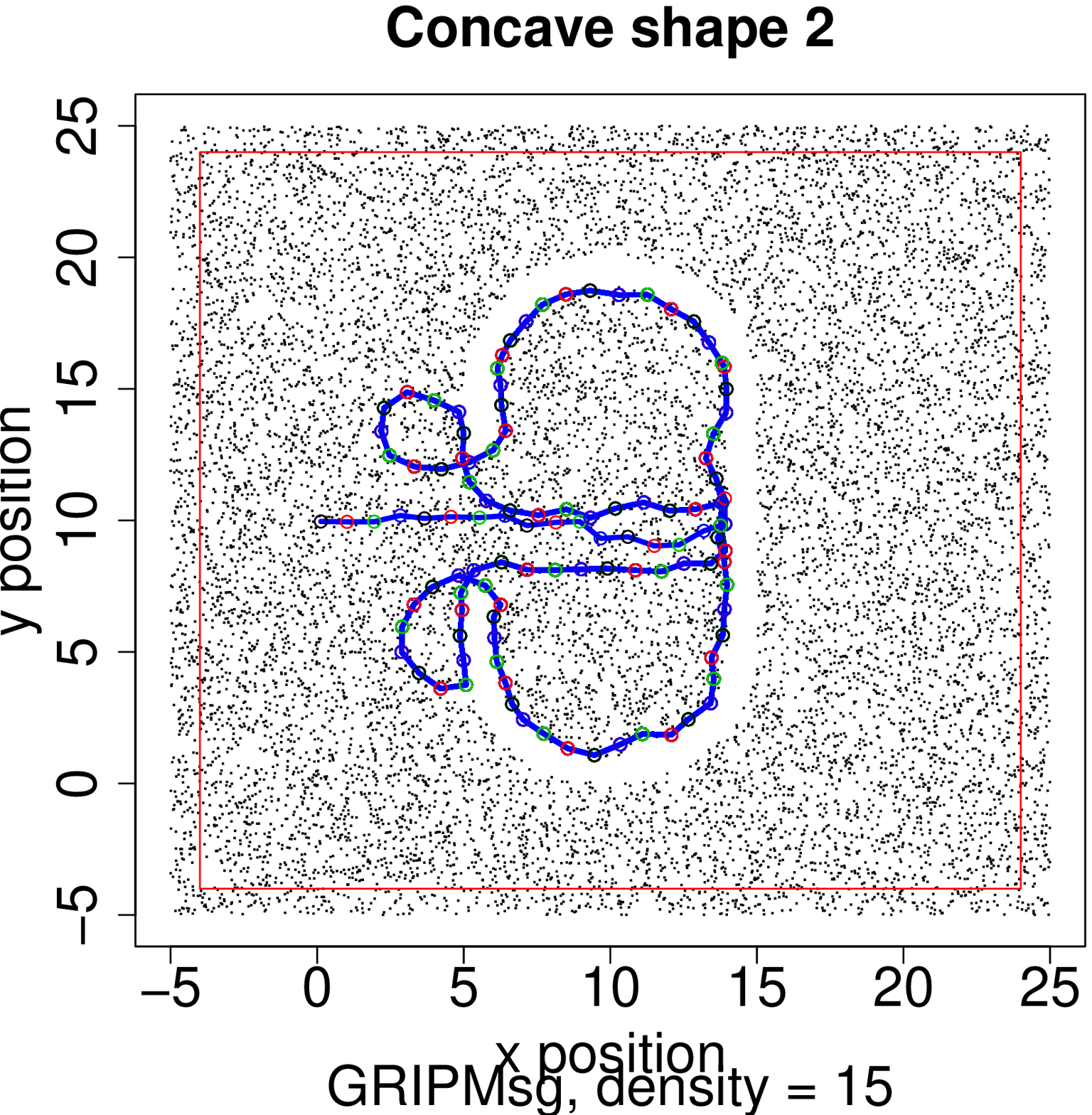}
}
\caption{Typical behavior of \GRIC (non-random version) for different obstacle shapes.}
\label{obstacles}
\label{paths}
\end{figure*}

Our second finding is that \GRIC is successful at efficiently (using
short paths) avoiding large convex obstacles like in figure \ref{path stripe}, 
and even some trap shaped concave
obstacles like the one in figures \ref{path U} and \ref{path cuve good}.  
Since to our knowledge all previous lightweight algorithms with no complicated
topology discovery and not relying on a planarization phase fail for
all but the simplest obstacles (simpler than all those considered in
this work), we conclude that \GRIC outperforms all
algorithms from that category.
As obstacles become increasingly hard (i.e. as they are more and more
shaped like a message trap), the density required for high success rate
increases, as can be seen from figures \ref{fig simulation results}.
There is a limit to the type of obstacles \GRIC can avoid. (For
example, \GRIC will not get a message out of a maze). We identify a
trap shaped obstacles (figure \ref{path cuve bad}) for which \GRIC
performs poorly, as can be seen on figures \ref{metric cuve bad} and
\ref{metric cuve bad2}, in the sense that except for very high
densities, the success probability stays low and even when considering very
high densities, although success probability raises,  the path
length stays sub-optimal.

When comparing path lengths, we can observe that \GRIC is close to
greedy optimal. 
That is, besides being successful (i.e. delivering
messages to their destination), it is also very efficient in the sense
that the path travelled is very short.
Indeed, as can be seen in figure \ref{obstacles}, the \GRIC algorithm
proceeds greedily before reaching the obstacle
(i.e. selecting nearest to destination next hop sensors thus
progressing as well as possible). As soon as the obstacle is reached,
the path taken follows quite closely the obstacle shape. Finally, when
the obstacle is bypassed, the path is more or less a direct line to
the destination. Thus, the performance of our algorithm compares very
well to what we can call a ``local knowledge'' optimal algorithm in
the case of no global knowledge of the network, i.e. no a priori knowledge of the presence of obstacles before reaching them.
\begin{figure*}[htb]
\centering
\subfigure[Success, no obstacle.]{
  \label{metric void}
  \includegraphics[angle=0,width=.48\textwidth]{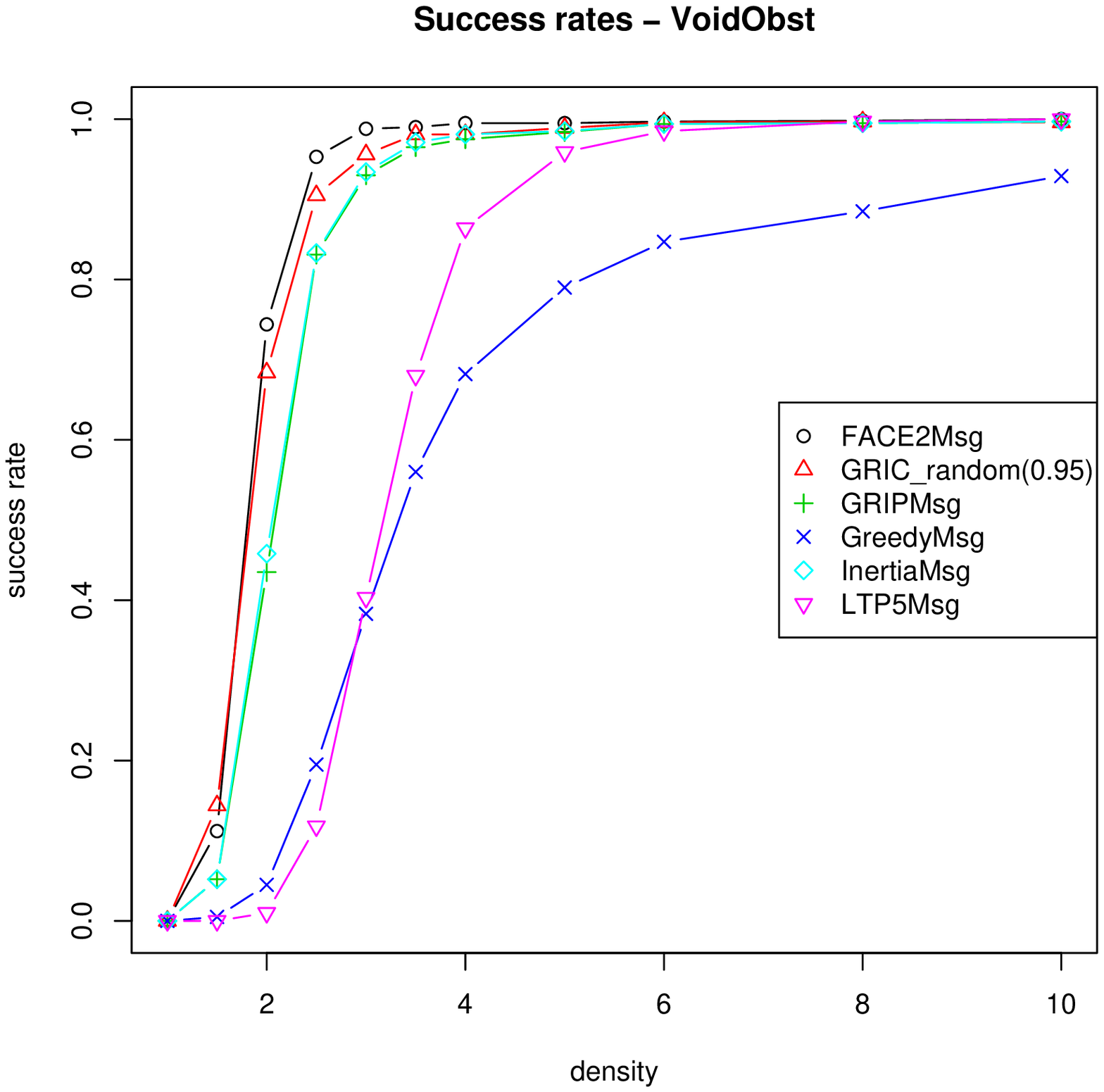}
}
\subfigure[Hops, no obstacle.]{
  \label{metric void2}
  \includegraphics[angle=0,width=.48\textwidth]{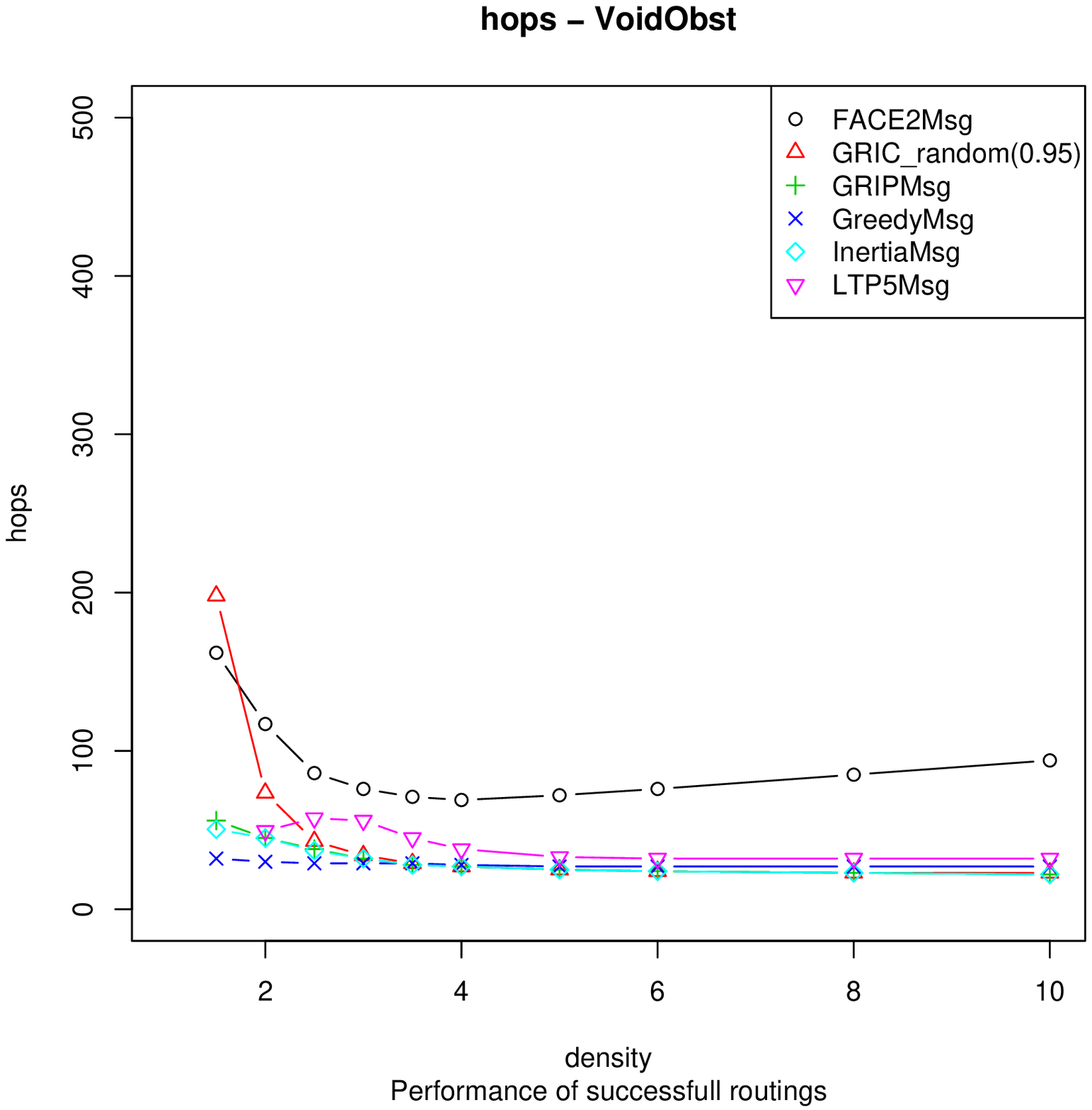}
}
\end{figure*}
\begin{figure*}[htb]
\centering
\subfigure[Success, stripe.]{
  \label{metric stripe}
  \includegraphics[angle=0,width=.48\textwidth]{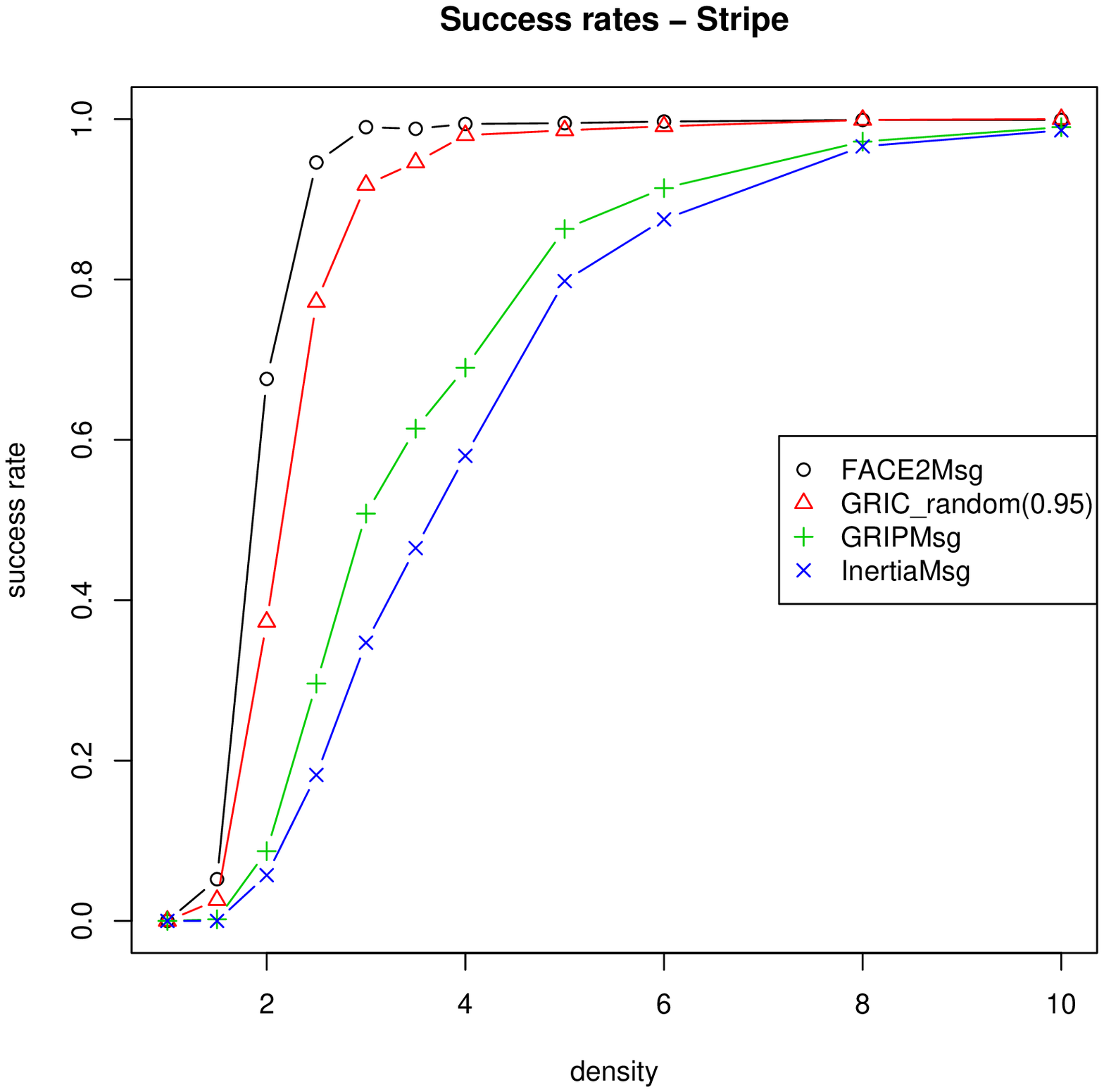}
}
\subfigure[Hops, stripe.]{
  \label{metric stripe2}
  \includegraphics[angle=0,width=.48\textwidth]{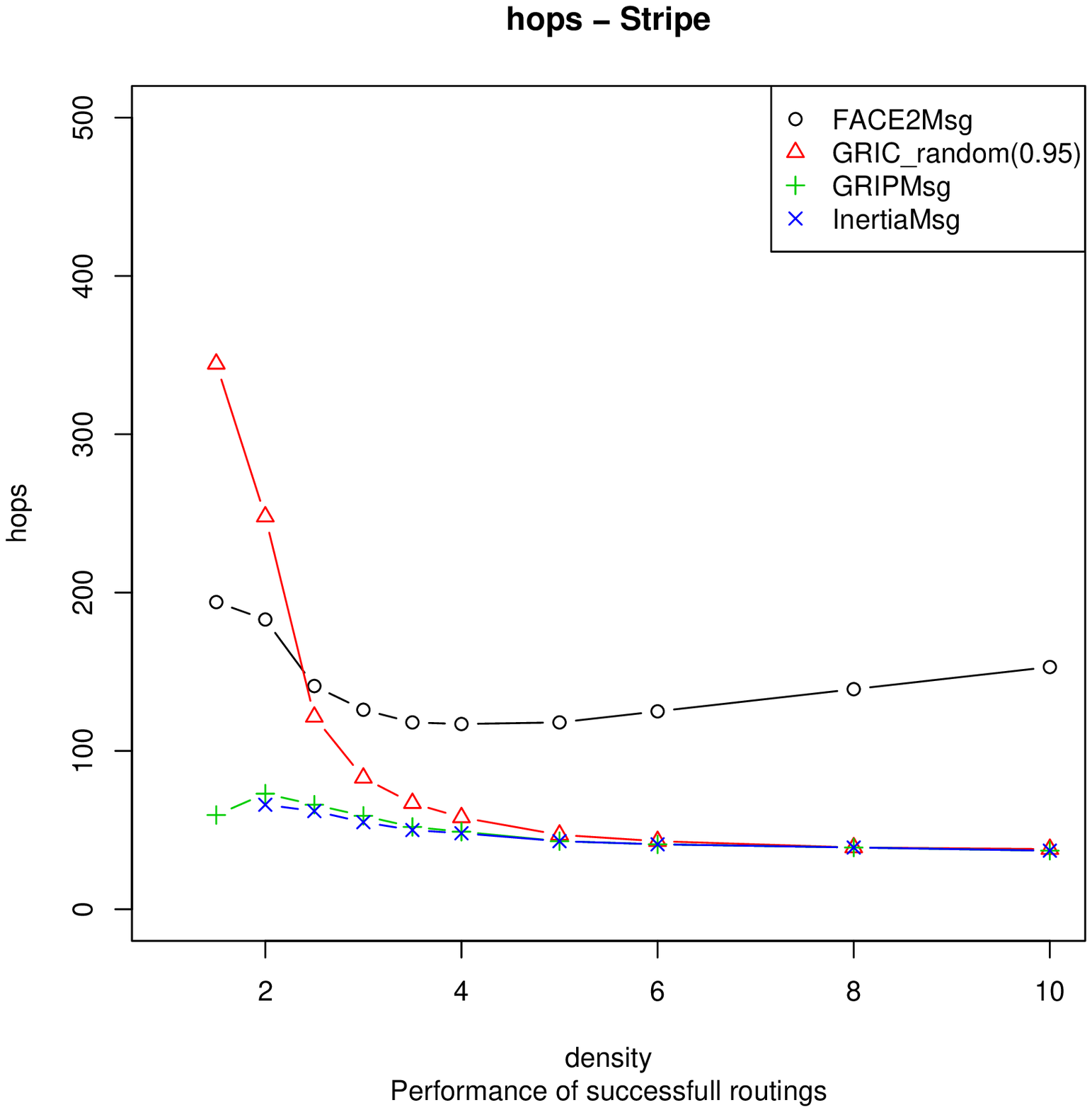}
}
\end{figure*}
\begin{figure*}[htb]
\centering
\subfigure[Success, U shape.]{
  \label{metric U}
  \includegraphics[angle=0,width=.48\textwidth]{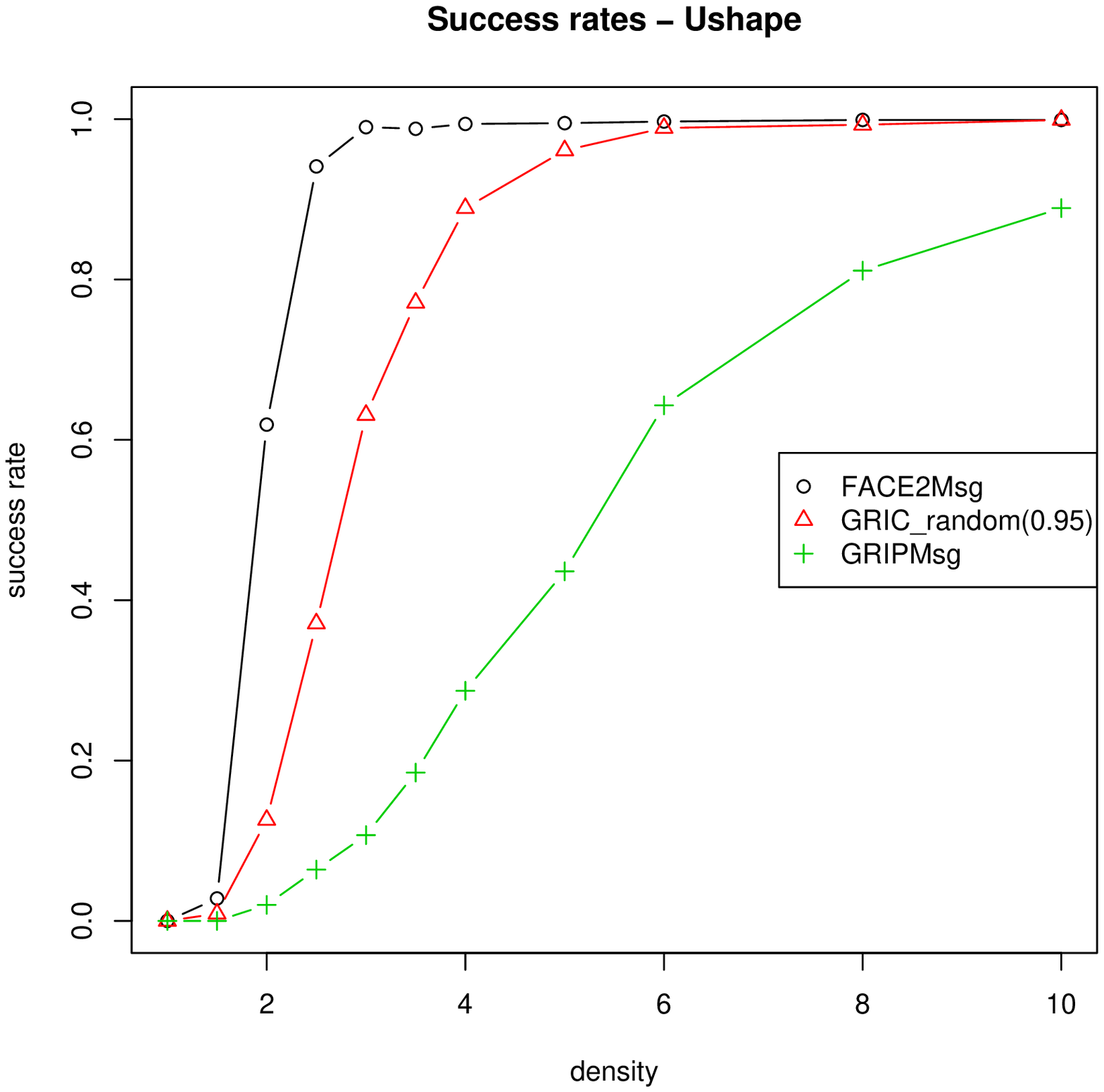}
}
\subfigure[Hops, U shape.]{
  \label{metric U2}
  \includegraphics[angle=0,width=.48\textwidth]{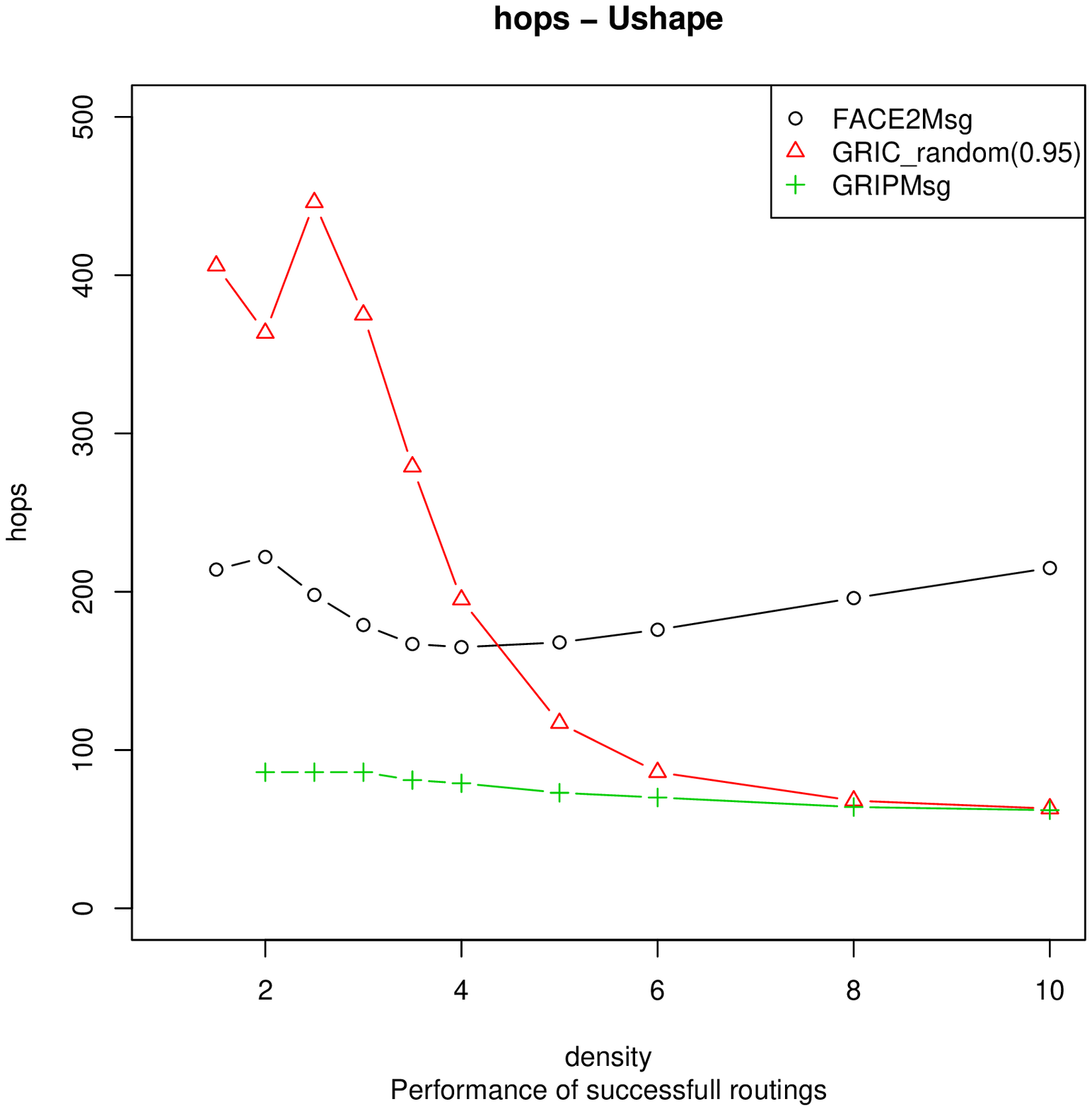}
}
\end{figure*}
\begin{figure*}[htb]
\centering
\subfigure[Success rates, concave shape 1.]{
  \label{good1}
  \includegraphics[angle=0,width=.48\textwidth]{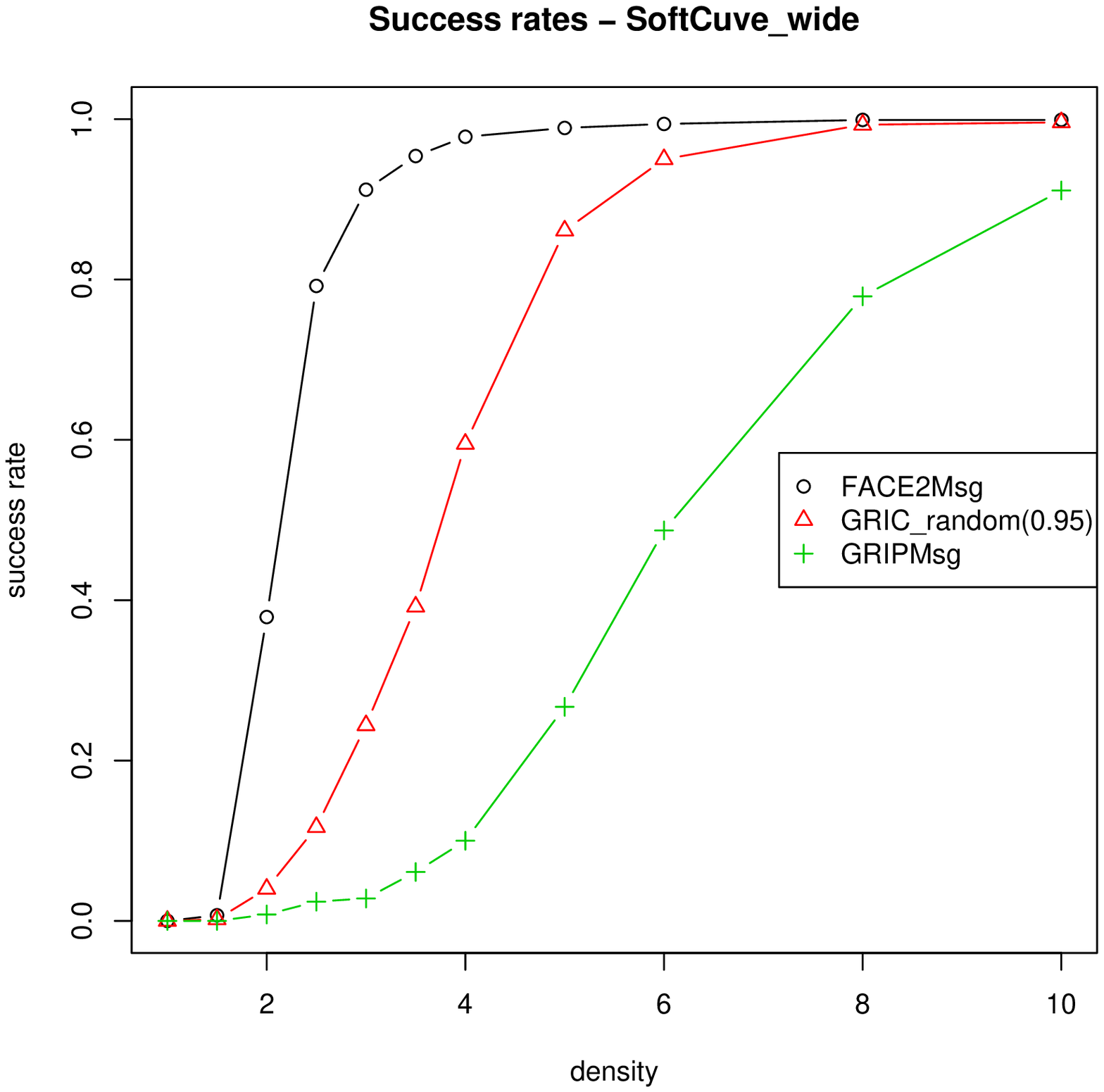}
}
\subfigure[Hops, concave shape 1.]{
  \label{good2}
  \includegraphics[angle=0,width=.48\textwidth]{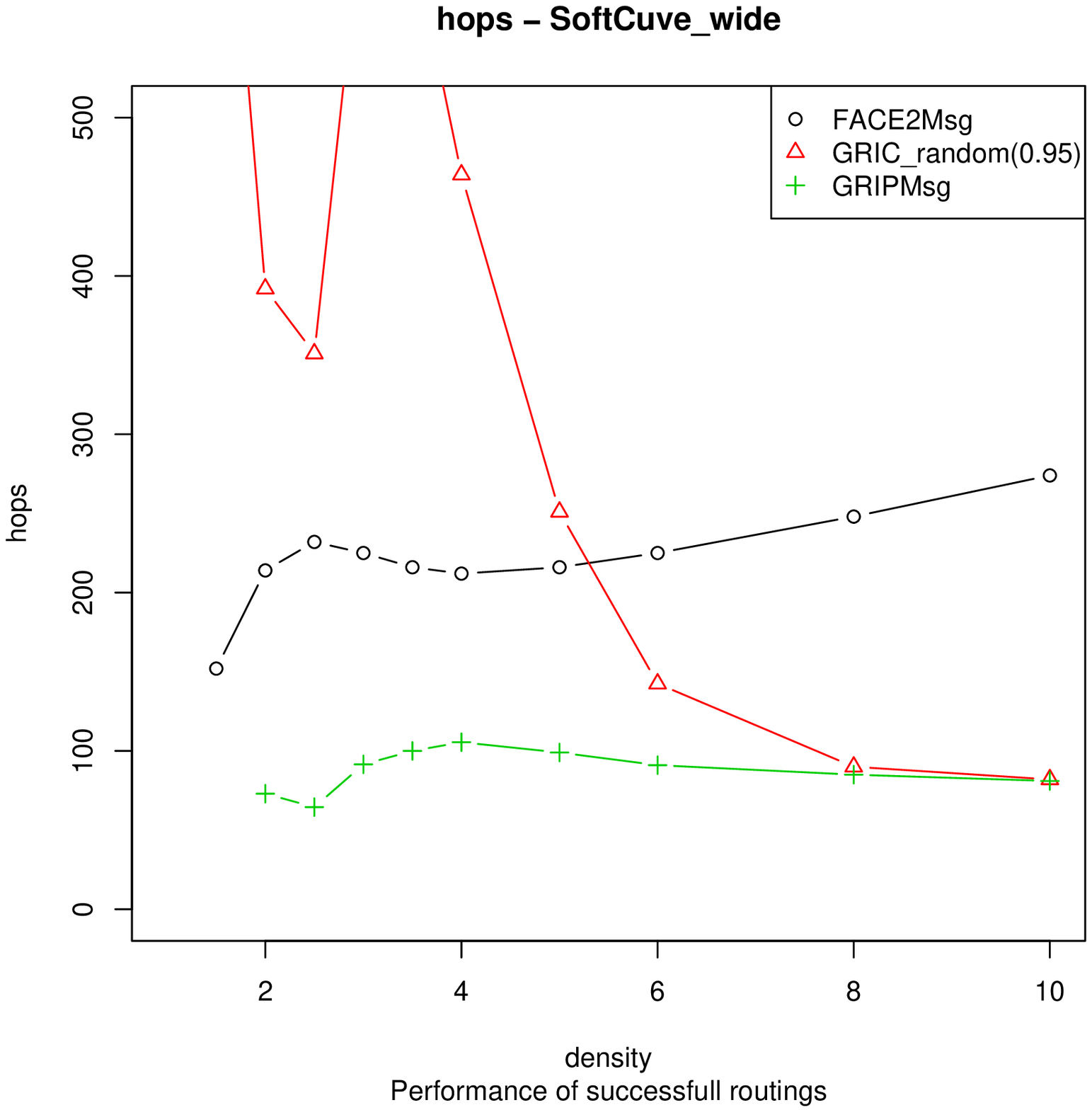}
}
\end{figure*}
\section{Detailed analysis}
\subsection{Routing holes}
Our first experimental results are in the case of no obstacles.
In this case, the only difficulty is routing messages around routing
holes following the local minimum problem, which is inherently
dependent on network density. The results are summarised in figure 
\ref{metric void}. We can see that for
medium densities ($d=5$), every protocol behaves well in terms of
success rates ($100\%$ success rates, except for greedy which has only 
$80\%$ success rate). As well as in terms of the distance and
hop-count metrics (the hop-counts are shown in figure \ref{fig simulation results} and the plot of travelled distances will be found in the \ref{appendix figures} section
of this appendix.). However, when the density drops, algorithms start
failing: first \textbf{LTP} (when the density gets below $d=4$), then
\textbf{Inertia} and \GRICminus when density gets below $2.5$. We observe
that although success rates drop for $d\leq 2$ for \GRICplus, so does
the success rate of \textbf{FACE2}, which we interpret as the fact that \GRICplus fails only
when the network is disconnected. In terms of success rates, this
means that \textbf{Inertia} and \GRICminus are quite good, but that \GRICplus is
even better: the power of the randomness component comes into play and
makes \GRICplus superior to the other protocols.
Looking at the distance and hop-count, \GRICplus is almost as good as \GRICminus
and \textbf{Inertia} as soon as $d=2.5$. For $d<2.5$ \GRICplus is not as
good but this is because \textbf{Inertia} and \GRIC fail on hard
instances (i.e. they have lower success rates). 

Summarizing, \GRICplus successfully (with success rate
close to 100\% if the netowrk is connected) and efficiently routes
messages in localised networks, even in the case of \emph{extremely low
densities}.

\subsection{Obstacles}
We next consider the case of a sensor network with obstacles.
We consider four different obstacle shapes: a stripe, a U shape, and
two types of semi-closed concave shapes. Those shapes can be seen on
figure \ref{obstacles} together with a typical path followed by a
message routed by our \GRIC algorithm (as can be seen, and we shall
explain why, the second type of concave shape makes \GRIC fail with
high probability).

We start by looking at the \emph{stripe} obstacle, which is a large
and hard to bypass obstacle. On figure
\ref{metric stripe} we see that \GRICplus successfully routes messages
around stripe shaped obstacles with high probability (around 95\%) even for
low densities ($d\simeq 3$) and almost certainly as soon as the
density is medium ($d=4$). Also, figure \ref{metric stripe} shows that the routing
is efficient since the path lengths are kept small. From our intuitive
understanding of the \GRICplus algorithm, we believe that this excellent
behavior will be reproduced for any convex shaped reasonable obstacle.

We next turn to concave shapes. The large concave shapes we consider are
hard obstacle instances, with shapes which can be thought of as being
message traps. However, such shapes may be reasonable in real world
scenarios. For example, a block of buildings could resemble the
U-shaped obstacles. The first concave shape may be less
realistic (it seems to be intentionally designed as a message trap),
however the \GRIC algorithm still behaves reasonably well (although not as
well as for concave shapes). The behavior of \GRICplus is very similar in
the case of the U shape obstacle and the first concave shape,
(c.f. figure \ref{metric U} and \ref{metric cuve good}). The conclusion is
that \GRICplus routes messages almost surely and efficiently around those
two obstacles, but only if the network density if medium to high ($d$
between $5$ and $6$). In short, the finding of those experiments is
that \GRICplus bypasses some concave shapes but only if the network
density is increased.
Finally, we show a concave shaped obstacle, 
the second concave shaped obstacle (c.f. figure
\ref{obstacles}), for which \GRICplus fails.
Simulation results, summarized on figure \ref{Simulation results for the concave shape 1},
show that the the success rate only becomes good for very high
densities ($d\simeq 8$). Furthermore, the path length is high (as can
be seen on plot of figure \ref{metric cuve bad} and \ref{metric cuve bad2}).
As a conclusion, the \GRICplus algorithm is incapable of efficiently
routing messages around the second concave shape of figure
\ref{obstacles}.
The reason the algorithm fails is the following.
We can see on the last plot of figure \ref{obstacles} what happens
with \GRICminus and the second concave shape. \GRICminus successfully routes the
message out of the obstacle. However, because following the contour of
the obstacle would require a very sharp turn, inertia prevents the
right-hand rule from making the message follow the contour of the
obstacle closely. This is also true for the other obstacles. The
difference here is that the influence of the message's destination on
the massage's trajectory unfortunately makes the message fall back
into the obstacle. The randomized version \GRICplus will eventually get
the message out of the obstacle after several repeated failed
attempts, but by then the message path, as seen on figure \ref{metric cuve bad}, 
will not be of competitive length anymore.
Finally, we present in figures \ref{good1} and \ref{good2} the results
for the concave shape 1 
(c.f. figure \ref{path cuve good}) which are comparable to those for the U
shape obstacle (c.f. figure \ref{path U}).
\subsubsection{Additional results}
\begin{figure*}[htb]
\centering
\subfigure[Concave shape 2.]{
\label{distance cuve bad}
\includegraphics[angle=0,width=.48\textwidth]{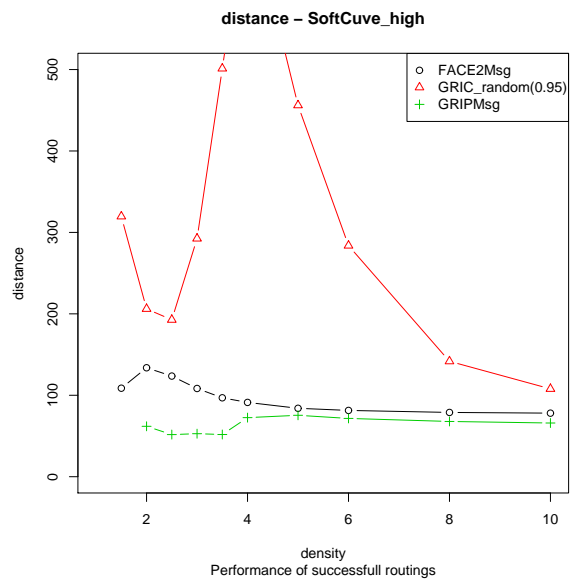}
}
\caption{Euclidean travelled distances}
\label{fig distances}
\label{metric cuve good}
\label{Simulation results for the concave shape 1}
\end{figure*}

This section contains additional figures.
Figure \ref{fig distances} gives the travelled median Euclidean
distances as measured in our experiments. It turns out that those
figures do not give out much more information than what is already
available through observing the hop-count metric given in figure
\ref{fig simulation results}, but we include them for completeness.
\label{appendix figures}
\begin{figure*}[htb]
\centering
\subfigure[No obstacle.]{
\label{distance void}
\includegraphics[angle=0,width=.48\textwidth]{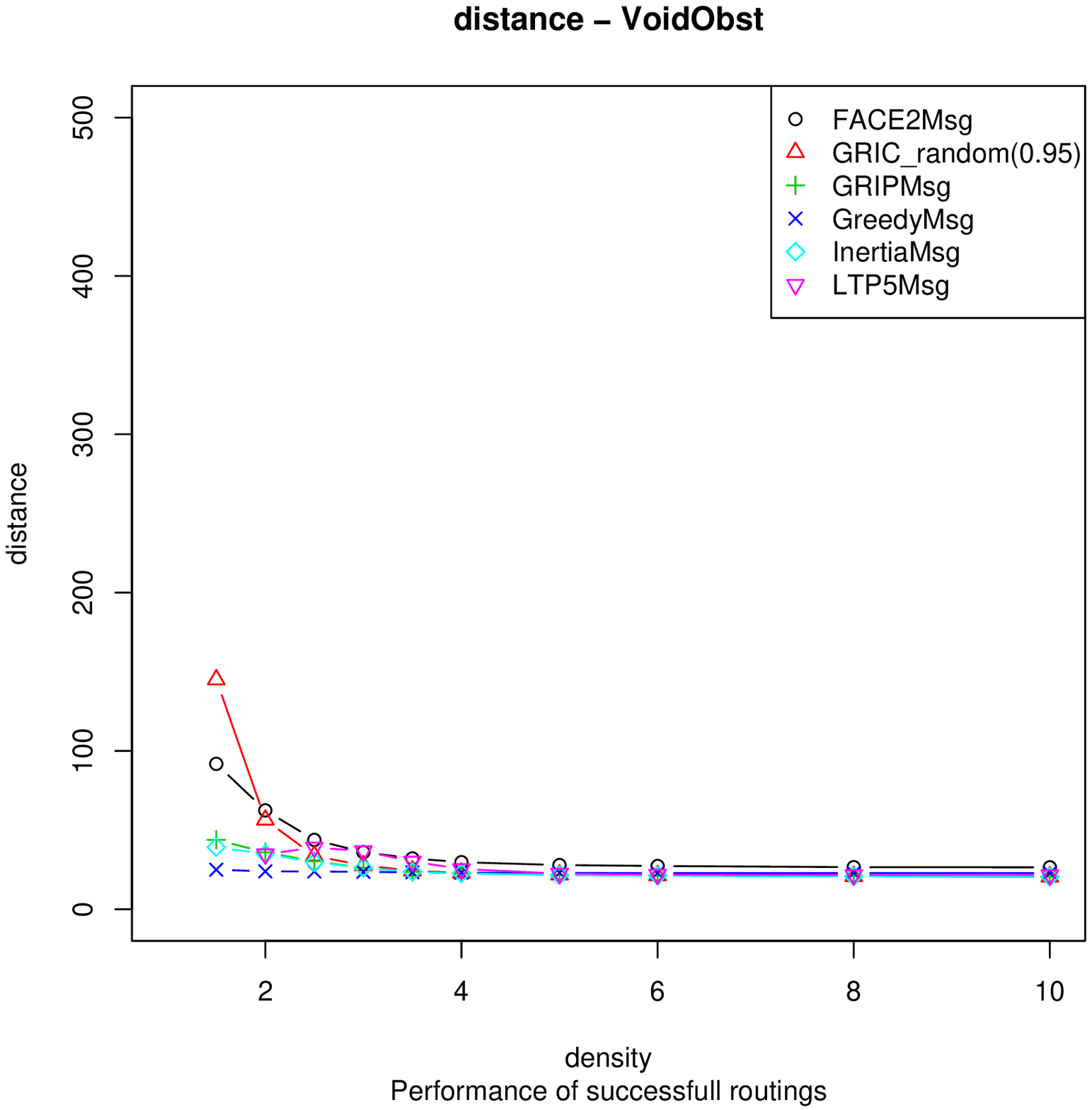}
}
\subfigure[Stripe.]{
\label{distance stripe}
\includegraphics[angle=0,width=.48\textwidth]{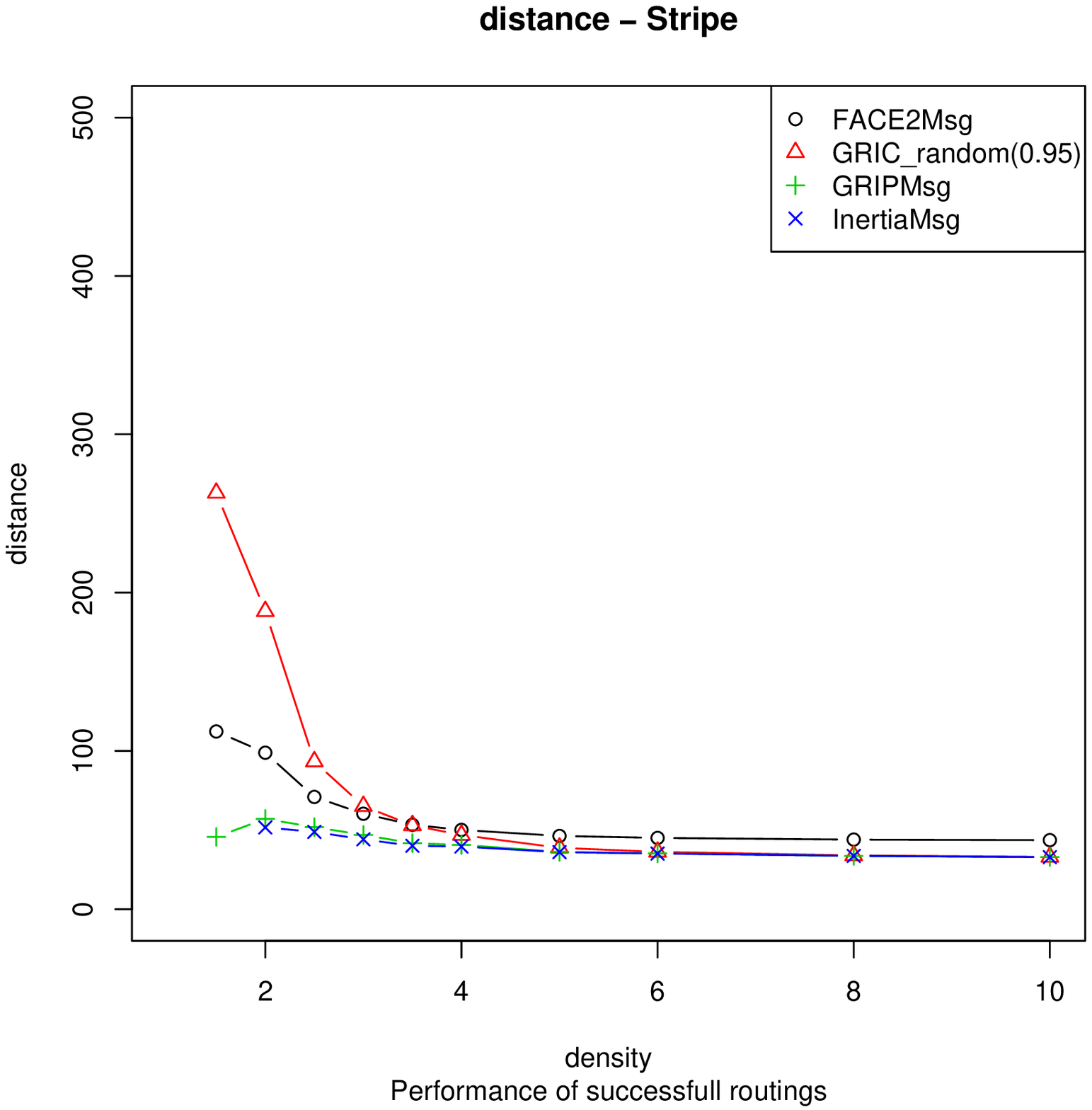}
}
\end{figure*}
\begin{figure*}[htb]
\centering
\subfigure[U shape.]{
\label{distance U}
\includegraphics[angle=0,width=.48\textwidth]{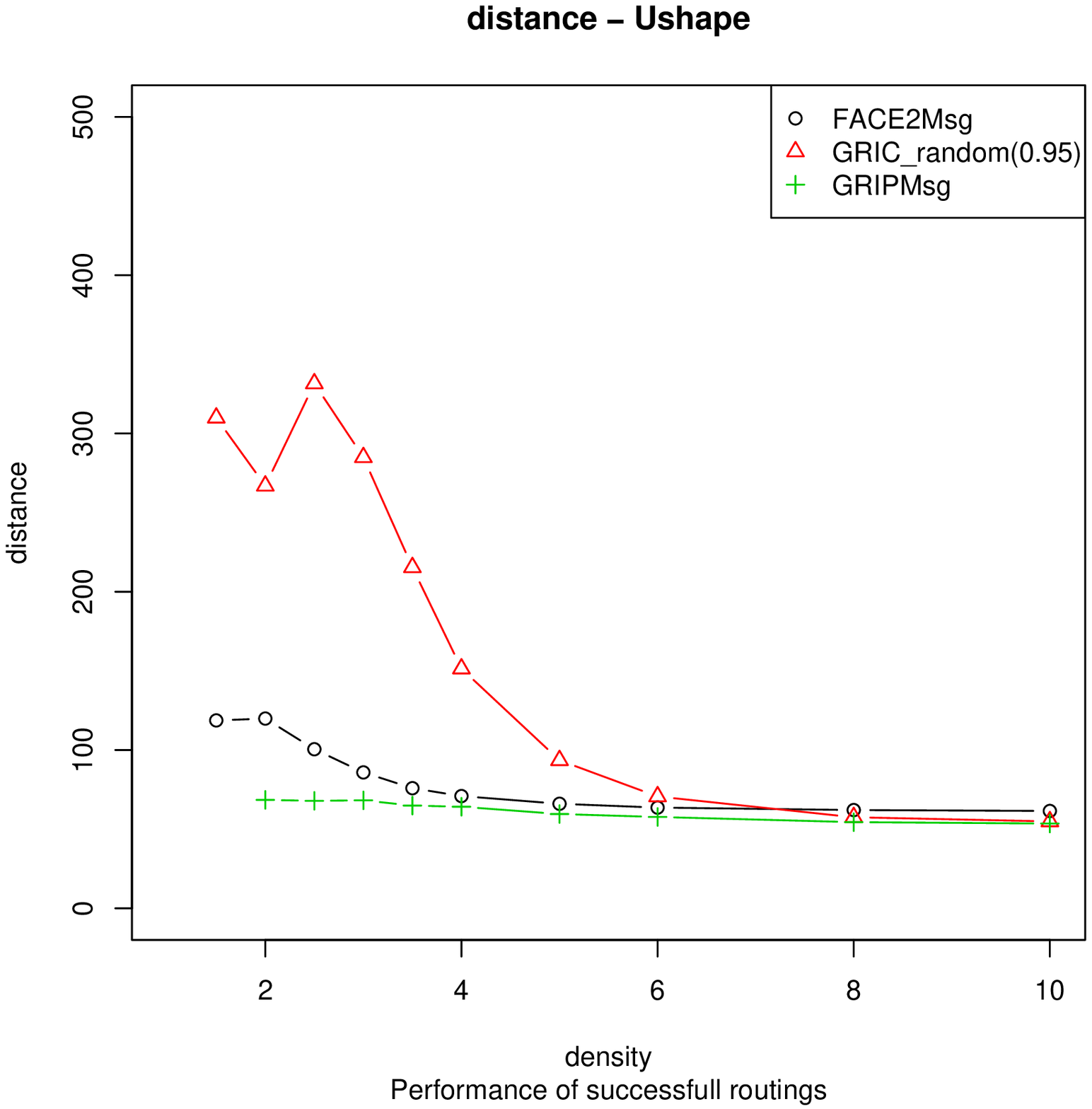}
}
\subfigure[Concave shape 1]{
\label{distance cuve}
\includegraphics[angle=0,width=.48\textwidth]{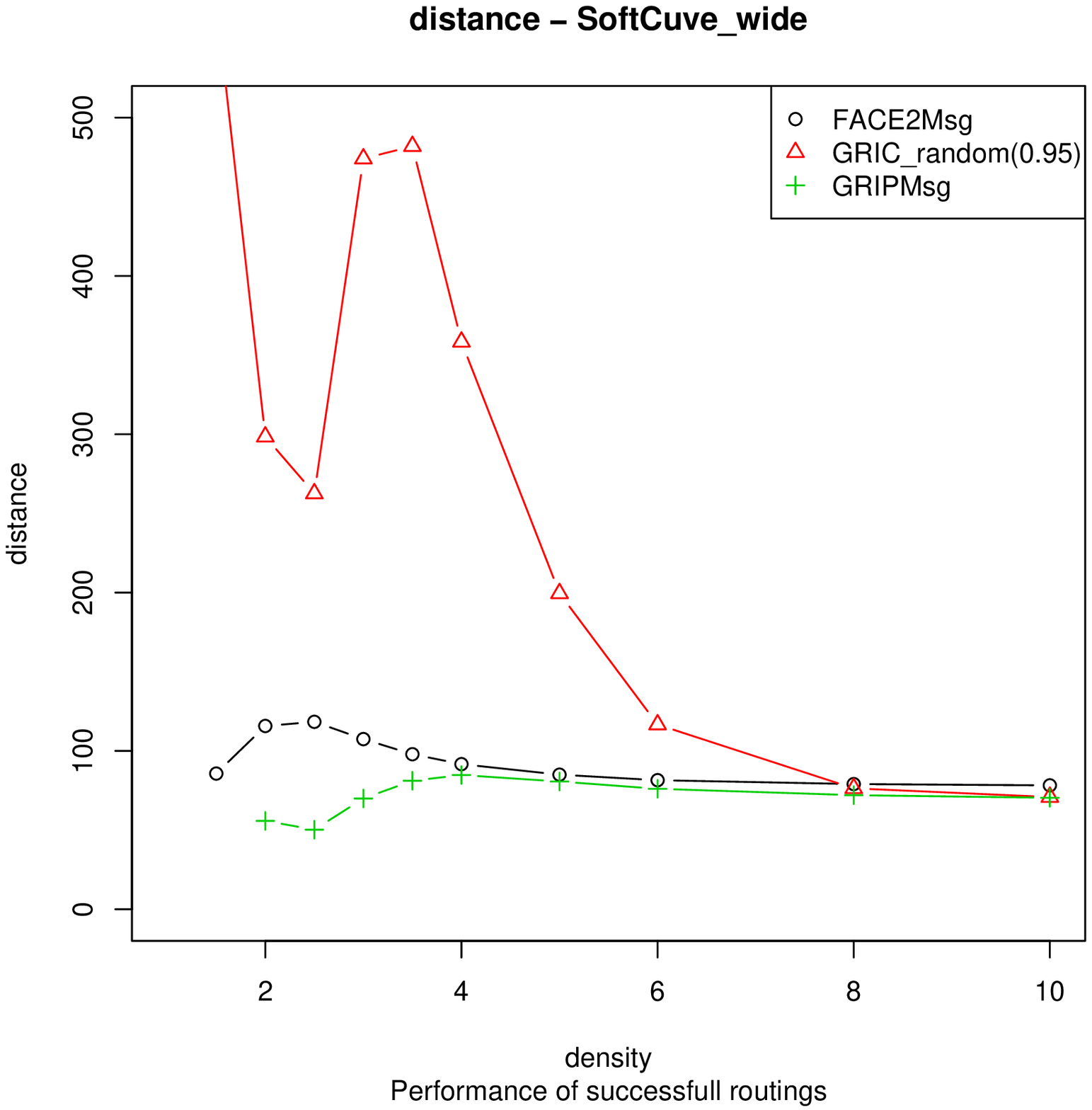}
}
\end{figure*}
\chapter{Conclusion}
We propose the \GRIC algorithm,  a new geographic routing algorithm for wireless
sensor networks. To our knowledge, it is the first successful
implementation of the right-hand rule idea \emph{-an idea
successfully used in the celebrated FACE family of algorithms-} that
runs on the full communication graph, i.e. that does rely
on a preliminary planarization phase. Therefore, it overcomes the limitations
inherent to the FACE family of algorithms while keeping its main
characteristics: near 100\% success rates (when no obstacles are
present) and the capacity to route messages around some complex
obstacles. Furthermore, routing is \emph{on demand} (i.e. it is
all-to-all) and requires no topology maintenance besides keeping track
of the outbound links, which do not even need to be symmetric.
\GRIC thus surpasses all previous similar ``lightweight'' and simple
geographic routing algorithms (which have comparatively low success
rates and simply fail in the presence of large obstacles). We
therefore believe that \GRIC substantially improves the state of the
art of geographic routing algorithms, arguably one of the most
important routing paradigms for wireless sensor networks.
\bibliographystyle{alpha}
\bibliography{georouting}		
\end{document}